\documentclass{amsart}
\usepackage{url}
\newtheorem{theorem}{Theorem}[section]
\newtheorem{lemma}[theorem]{Lemma}

\newtheorem{protocol}[theorem]{Protocol}

\theoremstyle{definition}
\newtheorem{definition}[theorem]{Definition}

\newtheorem{xca}[theorem]{Exercise}

\theoremstyle{remark}

\numberwithin{equation}{section}

\begin{document}
\title{\bf A polynomial time algorithm for SAT}
\author{Sergey Gubin }
\date{}

\begin{abstract}
Article presents the compatibility matrix method and illustrates it with the application to P vs NP problem. The method is a generalization of descriptive geometry: in the method, we draft problems and solve them utilizing the image creation technique. The method reveals: P = NP = PSPACE $\subseteq$ P/poly, etc.
\end{abstract}

\maketitle

\section*{Introduction}

This article presents the \emph{compatibility matrix method} \cite{01,02,03,04} and some of the results reported by the author at The Midwest Conference on Combinatorics, Cryptography and Computing (MCCCC) \cite{05,06,07}.
\newline\indent
The compatibility matrix method is a generalization of descriptive geometry on the combinatorial problems: the problems are "drafted" at first and then solved with the  image creation technique. We prove: the blueprints' size is polynomial in the problems' NTIME, and the image creation technique's computational complexity is polynomial in the blueprints' size. And we apply these results to the P vs NP problem \cite{08}.

\section{Compatibility matrix}
\label{s:cm}

The following definition formalizes the notion of problem's blueprints.
\begin{definition}
\label{d:1ten}
\emph{Compatibility matrix} $B$ is a symmetric Boolean box matrix with the diagonal Boolean matrices on its major diagonal:  
\begin{equation}
\label{e:1ten}
\left\{\begin{array}{rlcl}
1). & B& = & (B_{ij})_{n\times n} \\
2). & B_{ij} & = & (b_{\mu\nu}^{ij})_{m_i\times m_j} \\
3). & b_{\mu\nu}^{ij} & \in & \{false,true\} \\
4). & b_{\mu\nu}^{ij} & = & b_{\nu\mu}^{ji} \\ 
5). & b_{\mu\nu}^{ij} & \Rightarrow & i\neq j ~\vee~\mu = \nu \\
\end{array}\right.
\end{equation}
- where numbers $n$, $m_1$, $m_2$, $\ldots$, $m_n$, and maximum
\begin{equation} 
\label{e:11max}
m = \max_{i\in[n]} m_i
\end{equation}
are the appropriate sizes of the compatibility matrix. Boxes $B_{ij}$ are \emph{compatibility boxes}. Those components $b_{\mu\nu}^{ij}$ which equal $true$ are $true$-components, and the rest components are $false$-components.
\newline\indent
\emph{Index function} of compatibility matrix $B$ is any one-meaning function\footnote{For any integer number $N\geq 1$, $[N] = \{1,2,\ldots,N\}$.}
\[
\phi: i \in [n]~\mapsto~\phi(i)\in[m_i]
\]
Each index function $\phi$ defines a $2$-dimensional sub-array of the compatibility matrix which is a \emph{grid of components} or just a \emph{grid}:
\[
\{b_{\phi(i)\phi(j)}^{ij}\}_{i, j\in[n]}
\]
This grid is a \emph{solution grid} if all its components are $true$-components, i.e. if the grid's index function $\phi$ satisfies the following functional equation:
\begin{equation}
\label{e:1grid}
b_{\phi(i)\phi(j)}^{ij} = true,~i,j\in[n]
\end{equation}
- solution grid in compatibility matrix $B$ is an orthogonal lattice of $true$-components, one component per compatibility box.
\newline\indent
Further, a $true$-component $b_{\mu_0\nu_0}^{i_0j_0} = true$ is a \emph{noise} if functional equation \ref{e:1grid} is inconsistent subject to the following constrains:
\begin{equation}
\label{e:boundary}
\phi(i_0) = \mu_0,~\phi(j_0) = \nu_0
\end{equation}
Otherwise, $true$-component $b_{\mu_0\nu_0}^{i_0j_0} = true$ belongs to a solution grid.
\newline\indent
Furthermore, inversion of all noisy components in compatibility matrix \ref{e:1ten} will transform it into its \emph{general solution}\footnote{We could present any solution grid from compatibility matrix \ref{e:1ten} as a compatibility matrix of the same box structure as matrix $B$ and whose components all equal $false$ except those which belong to the grid. Then, the general solution would be the per-component disjunction of all those "particular solutions": $(a_{ij})\vee(b_{ij}) = (a_{ij}\vee b_{ij})$. The general solution does exist, is unique, and is a compatibility matrix on its own.}.
\end{definition}
\begin{xca}
\label{x:first}
The following Boolean box matrix is a compatibility matrix (as usual, values $true$ and $false$ are shown with 1 and 0, appropriately):
\[
B = \left(\begin{array}{cc}
B_{11} & B_{12} \\
B_{21} & B_{22} \\
\end{array}\right ) = \left(\begin{array}{cc|c}
1 & 0 & 1 \\
0 & 0 & 1 \\
\hline
1 & 1 & 1 \\
\end{array}\right)
\]
In this compatibility matrix, $n = 2$, $m = m_1 = 2$, and $m_2 = 1$. So, this compatibility matrix has two index functions ($\phi$ and $\psi$) and two grids of components, appropriately:
\[
\begin{array}{c||c}
\phi(1) = 1, ~\phi(2) = 1 & \psi(1) = 2, ~\psi(2) = 1 \\
\hline\hline
\left(\begin{array}{cc|c}
1 & . & 1 \\
. & . & . \\
\hline
1 & . & 1 \\
\end{array}\right) &
\left(\begin{array}{cc|c}
. & . & . \\
. & 0 & 1 \\
\hline
. & 1 & 1 \\
\end{array}\right) \\
\end{array}
\]
Index function $\phi$ creates a solution grid. Components $b_{21}^{12}$ and $b_{12}^{21}$ are noise, and inversion of them transforms this matrix into its general solution:
\[
G = \left(\begin{array}{cc|c}
1 & 0 & 1 \\
0 & 0 & 0 \\
\hline
1 & 0 & 1 \\
\end{array}\right) =
\left(\begin{array}{cc|c}
1 & 0 & 1 \\
0 & 0 & 0 \\
\hline
1 & 0 & 1 \\
\end{array}\right) ~\vee ~
\left(\begin{array}{cc|c}
0 & 0 & 0 \\
0 & 0 & 0 \\
\hline
0 & 0 & 0 \\
\end{array}\right)
\]
- it may be seen as the disjunction of all solution grids of matrix $B$.
\end{xca}
\begin{xca}
\label{x:222}
The following Boolean box matrix is another compatibility matrix:
\[
B = \left ( \begin{array}{cc|cc|cc}
1 & 0 & 1 & 0 & 0 & 1 \\
0 & 1 & 0 & 1 & 1 & 0 \\
\hline 
1 & 0 & 1 & 0 & 1 & 0 \\
0 & 1 & 0 & 1 & 0 & 1 \\
\hline 
0 & 1 & 1 & 0 & 1 & 0 \\
1 & 0 & 0 & 1 & 0 & 1 \\
\end{array}\right )
\]
In this compatibility matrix, $n = 3$ and $m = m_1 = m_2 = m_3 = 2$. Then, there are $8$ grids of components in this matrix. Yet, none of them is a solution grid. Then, this matrix's general solution equals $((0)_{2\times 2})_{3\times 3}$.
\end{xca}
Compatibility matrices are used to encode the natural problems and, sometimes, the whole families of the problems. There are not any requirements to the \emph{compatibility-matrix encoding} except the encoding's \emph{adequacy}:
\begin{protocol}
\label{p:post}
There are solution grids in the given compatibility matrix iff the encoded problem is consistent, i.e. iff it has solutions.
\end{protocol} 
Several examples of the adequate compatibility-matrix encoding are presented in Section \ref{s:x}. Also, Section \ref{s:x} specifies domain of the encoding.
\newline\indent
Due to Protocol \ref{p:post}, solution-grids in the compatibility-matrix encoding may be identified with the solutions of the encoded problem in the "topological sense." Then, the compatibility boxes are the solution's 2D-views; the compatibility matrix itself is an orthographic projection layout display; and functional equation \ref{e:1grid} is the image creation technique. 
\newline\indent
The compatibility-matrix encoding of the decision and search problems is appropriately the Post and Turing reduction of the problems to the solution grids existence problem (to find a solution grid we may: iterate the compatibility boxes and the $true$-components in the current box; nil all components in the current box and its transpose box except the current component and its transpose component, and test the resulting matrix on the solution grids existence; we move to the next component if the current component is noise, else we move to the next compatibility box). 
\newline\indent
General solution of the given compatibility matrix is a compatibility matrix of the encoded problem on its own. And, due to its definition, it contains $true$-components iff the given compatibility matrix contains solution grids. General solution is the useful signal carried by the compatibility matrix. To compute the general solution, we need inverse the noisy components. Such a revitalization of the compatibility matrix is called \emph{depletion}.

\section{Solution grids and general solution}

Let's solve functional equation \ref{e:1grid} for the given compatibility matrix \ref{e:1ten} with the compatibility matrix method itself. For that, let's device such an adequate compatibility-matrix encoding of the equation's solutions which will allow its solution with the naive image creation technique. The encoding's idea is illustrated in the following exercise.
\begin{xca} 
\label{x:observ}[The major observation]
Let $true$-component $b_{\mu_0\nu_0}^{i_0 j_0} = true$ be a noise in compatibility matrix \ref{e:1ten}. And let $\phi$ be a partial index function:
\[
\phi: i\in s\subset [n] \mapsto \phi(i) \in [m_i]
\]
Let $i_0,j_0\in s$, and let $\phi$ satisfy "boundary conditions" \ref{e:boundary}, and let
\[
\bigwedge_{i,j\in \mbox{\scriptsize Dom}(\phi)} b_{\phi(i)\phi(j)}^{ij} = true
\]
Let's use random walk method and extrapolate partial function $\phi$ as far as possible preserving value $true$ of this conjunction.
\newline\indent
We iterate set $[n]-\mbox{Dom}(\phi)$. For the current $i_1\in [n]-\mbox{Dom}(\phi)$, we search set $[m_{i_1}]$ for any such $\mu_{i_1}\in[m_{i_1}]$ that
\[
b_{\mu_{i_1}\mu_{i_1}}^{i_1i_1}\wedge \bigwedge_{i\in\mbox{\scriptsize Dom}(\phi)} b_{\phi(i)\mu_{i_1}}^{i i_1} = true
\]
And, when such index $\mu_{i_1}$ is found, we extrapolate function $\phi$:
\[
\mbox{Dom}(\phi) = \mbox{Dom}(\phi)\cup\{i_1\},~\phi(i_1) = \mu_{i_1}
\] 
After that, we move to the next $i_1\in [n]-\mbox{Dom}(\phi)$.
\newline\indent
Because $b_{\mu_0\nu_0}^{i_0 j_0} = true$ is a noise, each of these random walks will stuck in a dead-end: we will find that for some current $i_1\in [n]-\mbox{Dom}(\phi) \neq \emptyset$
\[
\bigvee_{\mu\in[m_{i_1}]} ( b_{\mu\mu}^{i_1i_1}\wedge \bigwedge_{i\in\mbox{\scriptsize Dom}(\phi)} b_{\phi(i)\mu}^{i i_1} ) = false
\]
- we would extrapolate our partial index function $\phi$ to a total index function which satisfies functional equation \ref{e:1grid} what is impossible, otherwise.
\end{xca}
Exercise \ref{x:observ} shows that functional equation \ref{e:1grid} can be adequately encoded in some intrinsic semi-global characteristics of the figure depicted on compatibility matrix \ref{e:1ten}: global in $m$ but not in $n$. Then, the compatibility-matrix encoding of the solution grids will be something like Gr$\ddot{\mbox{o}}$bner bases.
\newline\indent
Let's select $k\in[n]$. And let $Q_n^k$ be the set of all $k$-combinations of $n$. Let's arbitrarily enumerate set $Q_n^k$:
\begin{equation}
\label{e:Q}
Q_n^k = \{q_1,q_2,\ldots,q_{\mbox{\scriptsize C}_n^k}\}
\end{equation}
And let's arbitrarily enumerate each set $q_\alpha\in Q_n^k$:
\begin{equation}
\label{e:q}
q_\alpha = \{i_1^\alpha, i_2^\alpha, \ldots, i_k^\alpha \} \subseteq [n],~\alpha\in[\mbox{C}_n^k]
\end{equation}
For each $k$-combination $q_\alpha\in Q_n^k$, let $P_\alpha$ be the following Cartesian product:
\[
P_\alpha = [m_{i_1^\alpha}]\times [m_{i_2^\alpha}]\times\ldots\times [m_{i_k^\alpha}]
\]
And let's arbitrarily enumerate each set $P_\alpha$:
\begin{equation}
\label{e:P}
P_\alpha = \{p_1^\alpha,p_2^\alpha,\ldots,p_{\prod_{\iota\in[k]} m_{i_\iota^\alpha}}^\alpha\},~\alpha\in[\mbox{C}_n^k]
\end{equation}
- where $p_\beta^\alpha = (\mu_1^\beta,\mu_2^\beta,\ldots,\mu_k^\beta)$ is a $k$-tuple from set $P_\alpha$, $\beta\in [\prod_{\iota\in[k]} m_{i_\iota^\alpha}]$.
\newline\indent
And now, let's compute the following compatibility matrix from the given compatibility matrix \ref{e:1ten}:
\begin{equation}
\label{e:rk}
R(k) = (R_{\alpha_1\alpha_2}(k))_{|Q_n^k|\times |Q_n^k|} = ((r_{\beta_1\beta_2}^{\alpha_1\alpha_2})_{|P_{\alpha_1}|\times |P_{\alpha_2}|})_{|Q_n^k|\times |Q_n^k|}
\end{equation}
- where the matrix's components are
\begin{equation}
\label{e:rab}
\begin{array}{rl}
r_{\beta_1\beta_2}^{\alpha_1\alpha_2} = & (\alpha_1\neq\alpha_2 \vee  \bigwedge_{(i,\mu),(j,\nu)\in\{(i_1^{\alpha_1},\mu_1^{\beta_1}),\ldots,(i_k^{\alpha_1},\mu_k^{\beta_1})\}} b_{\mu\nu}^{ij}) \\
\wedge & (\alpha_1 = \alpha_2 \vee \bigwedge_{i_{\iota_1}^{\alpha_1}=i_{\iota_2}^{\alpha_2}}(\mu_{\iota_1}^{\beta_1}=\mu_{\iota_2}^{\beta_2})) \\
\end{array}
\end{equation}
- where $b_{\mu\nu}^{ij}$ are the components of the given compatibility matrix \ref{e:1ten}. 
\begin{xca}
\label{x:222_1}
Let's compute compatibility matrix \ref{e:rk} for the compatibility matrix from Exercise \ref{x:222} and values $k=1,2,3$.
\newline\indent
For $k = 1$: $Q_3^1 = \{q_1 = (i_1^1 = 1),q_2=(i_1^2=2),q_3=(i_1^3=3)\}$; $P_1 = \{p_1=(\mu_1^1=1),p_2=(\mu_1^2=2)\}$, $P_2 = \{p_1=(\mu_1^1=1),p_2=(\mu_1^2=2)\}$, and $P_3 = \{p_1=(\mu_1^1=1),p_2=(\mu_1^2=2)\}$; and
\[
R(1) = \left(\begin{array}{cc|cc|cc}
1 & 0 & 1 & 1 & 1 & 1 \\
0 & 1 & 1 & 1 & 1 & 1 \\
\hline 
1 & 1 & 1 & 0 & 1 & 1 \\
1 & 1 & 0 & 1 & 1 & 1 \\
\hline 
1 & 1 & 1 & 1 & 1 & 0 \\
1 & 1 & 1 & 1 & 0 & 1 \\
\end{array}\right)
\]
For $k=2$: $Q_3^2 = \{q_1 = (i_1^1=1,i_2^1=2),q_2=(i_1^2=1,i_2^2=3),q_3=(i_1^3=2,i_2^3=3)\}$; $P_1 = P_2 = P_3 = \{p_1=(\mu_1^1=1,\mu_2^1=1),p_2=(\mu_1^2=1,\mu_2^2=2),p_3=(\mu_1^3=2,\mu_2^3=1),p_4=(\mu_1^4=2,\mu_2^4=2)\}$; and 
\[
R(2) = \left(\begin{array}{cccc|cccc|cccc}
1 & 0 & 0 & 0 & 1 & 1 & 0 & 0 & 1 & 1 & 0 & 0 \\
0 & 0 & 0 & 0 & 1 & 1 & 0 & 0 & 0 & 0 & 1 & 1 \\
0 & 0 & 0 & 0 & 0 & 0 & 1 & 1 & 1 & 1 & 0 & 0 \\
0 & 0 & 0 & 1 & 0 & 0 & 1 & 1 & 0 & 0 & 1 & 1 \\
\hline
1 & 1 & 0 & 0 & 0 & 0 & 0 & 0 & 1 & 0 & 1 & 0 \\
1 & 1 & 0 & 0 & 0 & 1 & 0 & 0 & 0 & 1 & 0 & 1 \\
0 & 0 & 1 & 1 & 0 & 0 & 1 & 0 & 1 & 0 & 1 & 0 \\
0 & 0 & 1 & 1 & 0 & 0 & 0 & 0 & 0 & 1 & 0 & 1 \\
\hline
1 & 0 & 1 & 0 & 1 & 0 & 1 & 0 & 1 & 0 & 0 & 0 \\
1 & 0 & 1 & 0 & 0 & 1 & 0 & 1 & 0 & 0 & 0 & 0 \\
0 & 1 & 0 & 1 & 1 & 0 & 1 & 0 & 0 & 0 & 0 & 0 \\
0 & 1 & 0 & 1 & 0 & 1 & 0 & 1 & 0 & 0 & 0 & 1 \\
\end{array}\right),
\]
And for $k=3$: $Q_3^3=\{q_1=(i_1^1=1,i_2^1=2,i_3^1=3)\}$; $P_1 = \{p_1=(\mu_1^1=1,\mu_2^1=1,\mu_3^1=1), p_2=(\mu_1^2=1,\mu_2^2=1,\mu_3^2=2),p_3=(\mu_1^3=1,\mu_2^3=2,\mu_3^3=1), p_4=(\mu_1^4=1,\mu_2^4=2,\mu_3^4=2), p_5=(\mu_1^5=2,\mu_2^5=1,\mu_3^5=1), p_6=(\mu_1^6=2,\mu_2^6=1,\mu_3^6=2), p_7=(\mu_1^7=2,\mu_2^7=2,\mu_3^7=1), p_8=(\mu_1^8=2,\mu_2^8=2,\mu_3^8=2)\};$ and $R(3)=(0)_{8 \times 8}$.
\end{xca}
Basically, when $k\geq 2$, the diagonal components \ref{e:rab} encode the components of the given compatibility matrix \ref{e:1ten} while the off-diagonal components \ref{e:rab} encode the box structure of the given compatibility matrix \ref{e:1ten}.
\begin{lemma}
\label{l:ae}
When $k \geq 2$, solution grids in compatibility matrix \ref{e:rk} and solution grids in compatibility matrix \ref{e:1ten} are in one-to-one relation.
\end{lemma}
\begin{proof}
Let $\phi$ be index function of a solution grid in compatibility matrix \ref{e:1ten}. Then, for any $k$, the following function $\psi$ is index function of a solution grid in compatibility matrix \ref{e:rk}:
\[
\psi: \alpha\in[|Q_n^k|] \mapsto \beta\in [|P_\alpha|]
\]
- where $\beta = \psi(\alpha)$ is such that $p_\beta^\alpha = (\phi(i_1^\alpha),\phi(i_2^\alpha),\ldots,\phi(i_k^\alpha))$. And, for any $k$, matching $\phi\rightarrow\psi$ of these index functions is an injection.
\newline\indent
Let $\psi$ be index function of a solution grid in compatibility matrix \ref{e:rk}. Let $\alpha\in[|Q_n^k|]$. Then, $\beta = \psi(\alpha) \in [|P_\alpha|]$. Let $q_\alpha = (i_1^\alpha,i_2^\alpha,\ldots,i_k^\alpha)\in Q_n^k$ and $p_\beta^\alpha = (\mu_1^\beta,\mu_2^\beta,\ldots,\mu_k^\beta)\in P_\alpha$. Due to functional equation \ref{e:1grid},
\[
r_{\psi(\alpha_1)\psi(\alpha_2)}^{\alpha_1\alpha_2}\equiv true
\]
Then, due to the second conjunct on the right side of equalities \ref{e:rab}, the following function $\phi$ is an one-meaning function, and it is an index function of compatibility matrix \ref{e:1ten}:
\[
\phi: i_\iota^\alpha\in [n] \mapsto \mu_\iota^\beta \in [m_{i_\iota}]
\]
- where $\alpha\in [\mbox{C}_n^k]$, $\beta = \psi(\alpha)$, and $\iota\in[k]$. Matching $\psi\rightarrow \phi$ is an injection. And, due to the first conjunct on the right side of equalities \ref{e:rab},
\[
\bigwedge_{q_\alpha\in Q_n^k}~\bigwedge_{i,j\in[q_\alpha]} b_{\phi(i)\phi(j)}^{ij} = true
\]
Then, when $k\geq 2$, index function $\phi$ is a solution of functional equation \ref{e:1grid} for compatibility matrix \ref{e:1ten}: $b_{\phi(i)\phi(j)}^{ij} \equiv true,~i,j\in[n]$.
\end{proof}
Due to Lemma \ref{l:ae}, compatibility matrix \ref{e:rk} is a compatibility-matrix encoding of the compatibility matrix \ref{e:1ten} when $k\geq 2$. Let's deplete compatibility matrix \ref{e:rk} with the following \emph{zero propagation method}: we loop through compatibility boxes $R_{\alpha_1\alpha_2(k)}$ ($\alpha_1,\alpha_2\in[\mbox{C}_n^k]$) of compatibility matrix $R(k)$ and iterate the rows and columns in the current box; when the current row/column completely consists of the $false$-components, we propagate that value $false$ on the whole matrix $R(k)$ in the direction of the current row/column; and we continue while matrix $R(k)$ gets finalized.
\begin{xca}
The zero propagation method transforms compatibility matrix $R(2)$ from Exercise \ref{x:222_1} into the following compatibility matrix:
\[
R\acute{~}(2) = \left(\begin{array}{cccc|cccc|cccc}
1 & 0 & 0 & 0 & 0 & 1 & 0 & 0 & 1 & 0 & 0 & 0 \\
0 & 0 & 0 & 0 & 0 & 0 & 0 & 0 & 0 & 0 & 0 & 0 \\
0 & 0 & 0 & 0 & 0 & 0 & 0 & 0 & 0 & 0 & 0 & 0 \\
0 & 0 & 0 & 1 & 0 & 0 & 1 & 0 & 0 & 0 & 0 & 1 \\
\hline
0 & 0 & 0 & 0 & 0 & 0 & 0 & 0 & 0 & 0 & 0 & 0 \\
1 & 0 & 0 & 0 & 0 & 1 & 0 & 0 & 0 & 0 & 0 & 1 \\
0 & 0 & 0 & 1 & 0 & 0 & 1 & 0 & 1 & 0 & 0 & 0 \\
0 & 0 & 0 & 0 & 0 & 0 & 0 & 0 & 0 & 0 & 0 & 0 \\
\hline
1 & 0 & 0 & 0 & 0 & 0 & 1 & 0 & 1 & 0 & 0 & 0 \\
0 & 0 & 0 & 0 & 0 & 0 & 0 & 0 & 0 & 0 & 0 & 0 \\
0 & 0 & 0 & 0 & 0 & 0 & 0 & 0 & 0 & 0 & 0 & 0 \\
0 & 0 & 0 & 1 & 0 & 1 & 0 & 0 & 0 & 0 & 0 & 1 \\
\end{array}\right)
\]
Let's notice, there still is noise in compatibility matrix $R\acute{~}(2)$.
\end{xca}
Basically, the zero propagation method realizes the light-ray notion. And, as in the descriptive geometry itself, it often fails to produce the general solution because of such geometrical features of the depicted figure as concaveness and alike. Nevertheless, we would like this method to work and transform compatibility matrix \ref{e:rk} into its general solution. 
\newline\indent
Let's restrict depletion in the zero propagation method to the diagonal compatibility boxes of matrix \ref{e:rk}. The restricted zero-propagation method can be expressed by the following system of material implications\footnote{The second group of these implications is present when $k<n$ only.}:
\begin{equation}
\label{e:mi1}
\left\{\begin{array}{rrcl}
1) & \chi_{\beta_1\beta_1}^{\alpha_1\alpha_1} & \Rightarrow & r_{\beta_1\beta_1}^{\alpha_1\alpha_1} \\
2) & \chi_{\beta_1\beta_1}^{\alpha_1\alpha_1} & \Rightarrow & \bigvee_{\{\beta|r_{\beta_1\beta}^{\alpha_1\alpha}=true\}} \chi_{\beta\beta}^{\alpha\alpha}, ~\alpha\in[\mbox{C}_n^k],~\alpha\neq\alpha_1 \\
3) & true & \Rightarrow & \bigvee_{\beta\in [P_{\alpha_1}]} \chi_{\beta\beta}^{\alpha_1\alpha_1} 
\end{array}\right.
\end{equation}
- where $\alpha_1\in [|Q_n^k]|] = [\mbox{C}_n^k]$, $\beta_1\in[|P_{\alpha_1}|]=[\prod_{i\in[k]} m_{i^{\alpha_1}}]$, and variables $\chi_{\beta_1\beta_1}^{\alpha_1\alpha_1}$ are the unknown final values of the appropriate diagonal components. 
\begin{xca}
\label{x:mi1}
(1) Replacement of all implications in system \ref{e:mi} by their CNF-expressions\footnote{$(x\Rightarrow y)~\Leftrightarrow~(\bar{x}\vee y)$} will transform system \ref{e:mi1} into a dual Horn SAT instance. (2) System \ref{e:mi1} is monotone: let $R_1(k)$ and $R_2(k)$ be two matrices \ref{e:rk}, and let\footnote{For two Boolean matrices $(a_{ij})$ and $(b_{ij})$, $(a_{ij})\Rightarrow(b_{ij})$ if $a_{ij}\Rightarrow b_{ij}$ for all $i$ and $j$. Material implication is a partial order on the set of all Boolean matrices.} $R_1(k)\Rightarrow R_2(k)$; then any solution of system \ref{e:mi1} for matrix $R_1(k)$ is a solution of the system for matrix $R_2(k)$. (3) Set of solutions of system \ref{e:mi1} is closed under disjunction\footnote{If true assignments 
\[
\chi_{\beta_1\beta_1}^{\alpha_1\alpha_1} = \eta_{\beta_1\beta_1}^{\alpha_1\alpha_1}\in\{false,true\} \mbox{~and~} \chi_{\beta_1\beta_1}^{\alpha_1\alpha_1}=\theta_{\beta_1\beta_1}^{\alpha_1\alpha_1}\in\{false,true\}
\]
are the solutions of system \ref{e:mi1}, then true assignment 
\[
\chi_{\beta_1\beta_1}^{\alpha_1\alpha_1}=\eta_{\beta_1\beta_1}^{\alpha_1\alpha_1}\vee \theta_{\beta_1\beta_1}^{\alpha_1\alpha_1}
\]
is a solution of system \ref{e:mi1} as well.}. (4) Material implication is a partial order on the set of all solutions of system \ref{e:mi1}. (5) If system \ref{e:mi1} is consistent (it has solutions, i.e. there are such true assignments to variables $\chi_{\beta_1\beta_1}^{\alpha_1\alpha_1}$ which satisfy all implications in the system), there is unique \emph{maximal solution} which equals the disjunction of all solutions. (6) And that maximal solution can be obtained with the unit propagation method\footnote{For dual Horn CNF, the unit propagation method is as follows: for every each negative single-literal clause $\neg c$ aka unit, we remove from the CNF all literals $c$ and all clauses containing literal $\neg c$; in this way,  we propagate the units while there are any; the CNF is unsatisfiable iff this propagation will eliminate all literals from a clause; otherwise, when the propagation halts, we assign value $false$ to all those variables which created the units, and we assign value $true$ to the rest variables.}.
\end{xca}
\emph{Minimal solution} of system \ref{e:mi1} is any true assignment $\lambda$, 
\begin{equation}
\label{e:lambda}
\chi_{\beta_1\beta_1}^{\alpha_1\alpha_1} = \lambda_{\beta_1\beta_1}^{\alpha_1\alpha_1} \in \{false,true\},
\end{equation}
which satisfies all implications in system \ref{e:mi1} and in which the number of values $true$ equals C$_n^k$:
\begin{equation}
\label{e:lambda-1}
\sum_{\lambda_{\beta_1\beta_1}^{\alpha_1\alpha_1}} 1 = \mbox{C}_n^k
\end{equation}
Due to the third group of implications in system \ref{e:mi1}, the minimal solutions of system \ref{e:mi1} are in one-to-one relation with the solution grids in compatibility matrix \ref{e:rk} ($\alpha_1\in[|Q_n^k|]=[\mbox{C}_n^k]$ and $\beta_1\in[|P_{\alpha_1}|]$):
\begin{equation}
\label{e:oneone}
\lambda_{\beta_1\beta_1}^{\alpha_1\alpha_1} = (\psi(\alpha_1)=\beta_1)  = \bigwedge_{\iota\in[k]}(\phi(i_\iota^{\alpha_1}) = \mu_\iota^{\beta_1}) 
\end{equation}
- where $\psi$ and $\phi$ are index functions of the appropriate solution grids in compatibility matrices \ref{e:rk} and \ref{e:1ten}. Therefore, we have one-one reduced the solution grids existence/search problem to the minimal solutions problem.
\newline\indent 
In the spirit of Ramsey theory, the following mini-max ($m$ is maximum \ref{e:11max}) is called the \emph{uniform threshold} for compatibility matrix \ref{e:rk}:
\begin{equation}
\label{e:ut1}
\kappa_1 = \min\{n, m+1\}
\end{equation}
\begin{lemma}
\label{l:mi1}
For any $k\geq \kappa_1$, any solution of system \ref{e:mi1} for compatibility matrix \ref{e:rk} is a disjunction the system's minimal solutions. Particularly, the system is consistent for $k\geq \kappa_1$ iff its minimal solutions do exist.
\end{lemma}
\begin{proof}
This lemma hods in the case when system \ref{e:mi1} is inconsistent because there are not any solutions of the system in this case. To prove this lemma for consistent system \ref{e:mi1}, let's use mathematical induction over $n$.
\newline\indent
For $n = k$, compatibility matrix \ref{e:rk} has only one compatibility box which is a diagonal Boolean matrix. So, this lemma obviously holds in this case. 
\newline\indent
Assuming that this lemma holds for some $n = l \geq k$, let's prove it for 
\[
n = l+1 > k
\]
Let $\sigma$ be our system \ref{e:mi1}, and let $\sigma\acute{~}$ be the following subsystem of $\sigma$:
\[
\sigma\acute{~}:~q_\alpha, q_{\alpha_1}\in Q_l^k
\]
System $\sigma\acute{~}$ is system \ref{e:mi1} for the $l\times l$ upper-left box corner $B\acute{~}$ of compatibility matrix \ref{e:1ten}. $B\acute{~}$ is a compatibility matrix on its own, and $n=l$ in  $B\acute{~}$. Therefore, this lemma holds for $\sigma\acute{~}$ due to our induction hypothesis. 
\newline\indent
Let $R\acute{~}$ be the submatrix of matrix \ref{e:rk} appropriate to subsystem $\sigma\acute{~}$. And let $\Lambda$ be the set of all minimal solutions of $\sigma\acute{~}$ (it is a finite set, $|\Lambda|\leq m^{kl}$):
\[
\Lambda = \{\lambda_1,\lambda_2,\ldots,\lambda_{|\Lambda|}\}
\] 
- where each $\lambda_{*\in[|\Lambda|]}$ is a true assignment \ref{e:lambda}/\ref{e:lambda-1} satisfying subsystem $\sigma\acute{~}$ ($\sum_{\lambda_{\beta_1\beta_1}^{\alpha_1\alpha_1}} 1 = \mbox{C}_l^k$). Then, due to our induction hypothesis, any solution $\chi_{\beta_1\beta_1}^{\alpha_1\alpha_1}=\eta_{\beta_1\beta_1}^{\alpha_1\alpha_1}$ of $\sigma\acute{~}$ (any true assignment $\chi_{\beta_1\beta_1}^{\alpha_1\alpha_1}=\eta_{\beta_1\beta_1}^{\alpha_1\alpha_1}$ which satisfies all implications in $\sigma\acute{~}$) is a disjunction of some true assignments from set $\Lambda$:
\[
\eta_{\beta_1\beta_1}^{\alpha_1\alpha_1} = \bigvee_{\lambda\in\eta\subseteq \Lambda} \lambda_{\beta_1\beta_1}^{\alpha_1\alpha_1}
\]
- where $\eta$ is the appropriate subset of $\Lambda$, appropriate to true assignment $\chi_{\beta_1\beta_1}^{\alpha_1\alpha_1}=\eta_{\beta_1\beta_1}^{\alpha_1\alpha_1}$. Let's notice, this decomposition is not unique, in general.
\newline\indent
Now, let the following true assignment be a solution of system $\sigma$:
\[
\chi_{\beta_1\beta_1}^{\alpha_1\alpha_1}=\theta_{\beta_1\beta_1}^{\alpha_1\alpha_1}\in\{false,true\},~\alpha_1\in[\mbox{C}_{l+1}^k]
\] 
Then, subset $q_{\alpha_1}\in Q_{l}^k$ of these assignments is a solution of subsystem $\sigma\acute{~}$. Then, there is subset $\theta\subseteq \Lambda$ such that
\[
\theta_{\beta_1\beta_1}^{\alpha_1\alpha_1} = \bigvee_{\lambda\in\theta} \lambda_{\beta_1\beta_1}^{\alpha_1\alpha_1},~q_{\alpha_1}\in Q_l^k
\]
Then, due to one-to-one relation \ref{e:oneone},
\[
\theta_{\beta_1\beta_1}^{\alpha_1\alpha_1} = \bigvee_{\lambda\in\theta} (\psi_\lambda(\alpha_1)=\beta_1) = \bigvee_{\lambda\in\theta} ~\bigwedge_{\iota\in[k]}(\phi_\lambda(i_\iota^{\alpha_1}) = \mu_\iota^{\beta_1}),~q_{\alpha_1}\in Q_l^k
\]
- where $\psi_\lambda$ and $\phi_\lambda$ are index functions \ref{e:oneone} of the solution grids in compatibility matrices $R\acute{~}$ and $B\acute{~}$ appropriate to the minimal solutions $\lambda\in\theta$ of subsystem $\sigma\acute{~}$ of system $\sigma$. Then, due to the second and first groups of implications in system $\sigma$ (where $n=l+1\geq 2$), for any minimal solution $\lambda\in \theta$, $k$-combination $q_{\alpha_1}\in Q_l^k$, and $k$-combination $q_\alpha\in Q_{l+1}^k - Q_l^k \neq \emptyset$
\[
true = \theta_{\psi_\lambda(\alpha_1)\psi_\lambda(\alpha_1)}^{\alpha_1\alpha_1}~\Rightarrow~ \bigvee_{\{\beta|r_{\psi_\lambda(\alpha_1)\beta}^{\alpha_1\alpha}=true\}} \theta_{\beta\beta}^{\alpha\alpha}~\Rightarrow~ \bigvee_{\{\beta|r_{\psi_\lambda(\alpha_1)\beta}^{\alpha_1\alpha}=true\}} r_{\beta\beta}^{\alpha\alpha}
\]
Then (see Exercise \ref{x:observ}), because $k\geq \kappa_1$, each partial index function $\phi_\lambda$, $\lambda\in\theta$, can be extrapolated from set $[l]$ on set $[l+1]$ for each $\theta_{\beta_1\beta_1}^{\alpha_1\alpha_1} = true$, $q_{\alpha_1}\in Q_{l+1}^k - Q_l^k$, due to equalities \ref{e:rab} and the rest implications in system $\sigma$. Each of these extrapolations satisfies functional equation \ref{e:1grid} for the given compatibility matrix \ref{e:1ten} and creates a minimal solution \ref{e:oneone} of $\sigma$. And solution $\chi_{\beta_1\beta_1}^{\alpha_1\alpha_1}=\theta_{\beta_1\beta_1}^{\alpha_1\alpha_1}$ of $\sigma$ is disjunction of all those minimal solutions. Therefore, this lemma holds for $n=l+1$. 
\end{proof}
System \ref{e:mi1} functions as follows: the first group of implications is an input; the third group of implications filters the input against a "rule" (the index wiring); and the third group of implications tests the result.
\begin{theorem}
\label{t:mi1}
For compatibility matrix \ref{e:1ten}:
\newline 
1). Let's compute compatibility matrix \ref{e:rk} for any $k\geq \kappa_1$;
\newline
2). Let's compile dual Horn CNF (it will be always satisfiable)
\[
\begin{array}{c}
h_2 = \bigwedge_{r_{\beta_1\beta_1}^{\alpha_1\alpha_1}}\neg \chi_{\beta_1\beta_1}^{\alpha_1\alpha_1}~\wedge~\bigwedge_{\alpha_1,\alpha\in[|Q_n^k|],\alpha\neq\alpha_1}(\neg\chi_{\beta_1\beta_1}^{\alpha_1\alpha_1}\vee\bigvee_{\{\beta|r_{\beta_1\beta}^{\alpha_1\alpha}\}} \chi_{\beta\beta}^{\alpha\alpha}) \\
\end{array}
\]
3). Let's deploy unit propagation method (see Exercise \ref{x:mi1}) to formula $h_2$ and compute the appropriate satisfying true assignment
\[
\chi_{\beta_1\beta_1}^{\alpha_1\alpha_1} = \eta_{\beta_1\beta_1}^{\alpha_1\alpha_1}\in\{false,true\}; \\
\]
4). Let's compile Horn CNF (it will be always satisfiable)
\[
\begin{array}{rl}
h_1 = & \bigwedge_{~\eta_{\beta_1\beta_1}^{\alpha_1\alpha_1}}~\bigwedge_{(i,\mu),(j,\nu)\in\{(i_1^{\alpha_1},\mu_1^{\beta_1}),\ldots,(i_k^{\alpha_1},\mu_k^{\beta_1})\}}~\xi_{\mu\nu}^{ij}\\
\wedge & \bigwedge_{\neg\eta_{\beta_1\beta_1}^{\alpha_1\alpha_1}}~\bigvee_{(i,\mu),(j,\nu)\in\{(i_1^{\alpha_1},\mu_1^{\beta_1}),\ldots,(i_k^{\alpha_1},\mu_k^{\beta_1})\}}\neg\xi_{\mu\nu}^{ij} \\
\end{array};
\]
5). Let's deploy unit propagation method\footnote{For Horn CNF, the unit propagation method is as follows: for every each positive single-literal clause $c$ aka unit, we remove from the CNF all literals $\neg c$ and all clauses containing literal $c$; in this way,  we propagate the units while there are any; the CNF is unsatisfiable iff this propagation will eliminate all literals from a clause; otherwise, when the propagation halts, we assign value $true$ to all those variables which created the units, and we assign value $false$ to the rest variables.} to formula $h_1$ and compute the appropriate satisfying true assignment
\[
\xi_{\mu\nu}^{ij} = \theta_{\mu\nu}^{ij} \in\{false,true\}.
\]
The general solution of compatibility matrix \ref{e:1ten} equals $((\theta_{\mu\nu}^{ij})_{m_i\times m_j})_{n\times n}$.
\end{theorem}
\begin{proof}
It is a direct consequence of Lemma \ref{l:ae}, Lemma \ref{l:mi1} (we just removed the third group of implications from system \ref{e:mi1}), and equalities \ref{e:rab}.
\end{proof}
As a mater of fact, Theorem \ref{t:mi1} is an analog of Gauss exclusions method, and it resolves the search problem. For the decision problem, we may remove backfeeds from  system \ref{e:mi1}. That reduced system is as follows:
\begin{equation}
\label{e:mi12}
\left\{\begin{array}{rrcl}
1) & \chi_{\beta_1\beta_1}^{\alpha_1\alpha_1} & \Rightarrow & r_{\beta_1\beta_1}^{\alpha_1\alpha_1} \\
2) & \chi_{\beta_1\beta_1}^{\alpha_1\alpha_1} & \Rightarrow & \bigvee_{\{\beta|r_{\beta_1\beta}^{\alpha_1\alpha}=true\}} \chi_{\beta\beta}^{\alpha\alpha}, ~\alpha\in[\mbox{C}_n^k],~\alpha > \alpha_1 \\
3) & true & \Rightarrow & \bigvee_{\beta\in [P_1]} \chi_{\beta\beta}^{\alpha_1\alpha_1} 
\end{array}\right.
\end{equation}
This system is consistent iff the following dual Horn CNF is satisfiable:
\[
h_3 = \bigwedge_{r_{\beta_1\beta_1}^{\alpha_1\alpha_1}}\neg \chi_{\beta_1\beta_1}^{\alpha_1\alpha_1}~\wedge~\bigwedge_{\alpha_1,\alpha\in[|Q_n^k|],\alpha > \alpha_1}(\neg\chi_{\beta_1\beta_1}^{\alpha_1\alpha_1}\vee\bigvee_{\{\beta|r_{\beta_1\beta}^{\alpha_1\alpha}\}} \chi_{\beta\beta}^{\alpha\alpha}) ~\wedge~\bigvee_{\beta\in [P_1]} \chi_{\beta\beta}^{\alpha_1\alpha_1} 
\]
Formula $h_3$ is satisfiable iff there are solution grids in compatibility matrix \ref{e:1ten}. And satisfiability of formula $h_3$ can be tested with the unit propagation method. Computational complexity of the unit propagation method is linear in the formula size which is $O(m^k(\mbox{C}_n^k)^2)$. And we can control the computational complexity to some extend with $k$, $\kappa_1\leq k \leq n$. The minimal value of the computational complexity is $O(m^{\kappa_1}(\mbox{C}_n^{\kappa_1})^2)$ when $k = \kappa_1$.
\newline\indent
Substitution $\chi_{\beta\beta}^{\alpha\alpha} = \neg \zeta_{\beta\beta}^{\alpha\alpha}$ transforms system \ref{e:mi12} into the following monotone circuit for the decision problems:
\begin{equation}
\label{e:mi13}
\left\{\begin{array}{rrcl}
1) & \zeta_{\beta_1\beta_1}^{\alpha_1\alpha_1} & \Leftarrow & \neg r_{\beta_1\beta_1}^{\alpha_1\alpha_1} \\
2) & \zeta_{\beta_1\beta_1}^{\alpha_1\alpha_1} & \Leftarrow & \bigvee_{\alpha\in[\mbox{\scriptsize C}_n^k],~\alpha > \alpha_1} \bigwedge_{\{\beta|r_{\beta_1\beta}^{\alpha_1\alpha}=true\}} \zeta_{\beta\beta}^{\alpha\alpha} \\
3) & \zeta_0 & \Leftarrow & \bigwedge_{\beta\in [P_1]} \zeta_{\beta\beta}^{\alpha_1\alpha_1} 
\end{array}\right.
\end{equation}
In this circuit, values $\neg r_{\beta_1\beta_1}^{\alpha_1\alpha_1}$ are an input, and value $\zeta_0$ is output. The output equals $false$ iff the compatibility-matrix encoded problem is consistent. The circuit's depth and width are $O(m^k\mbox{C}_n^k)$. The circuit's depths and width can be controlled with $k$, and their minimum is $O(m^{\kappa_1}\mbox{C}_n^{\kappa_1})$.
\newline\indent
Strictly speaking, circuit \ref{e:mi13} is not uniform. Yet, it can be fixed with the protocoling of enumerations \ref{e:Q}, \ref{e:q}, and \ref{e:P} for different box structures of compatibility matrix \ref{e:1ten}. And the number of the different box structures can be reduced with, for example, the padding of the smaller compatibility boxes by the rows/columns entirely filled with $false$. Then, systems \ref{e:mi1}, \ref{e:mi12}, and \ref{e:mi13} can be tabulated and wired. It would reduce the compatibility matrix method to the compatibility-matrix encoding.
\newline\indent
Another approach to the uniformity is to rid of the references to the components of  compatibility matrix \ref{e:rk} in the indices of system \ref{e:mi1} and its derivatives. It can be accomplished with, for example, the replacement of compatibility matrix \ref{e:rk} by the following compatibility matrix:
\begin{equation}
\label{e:sk}
S(k) = (S_{\alpha_1\alpha_2}(k))_{|Q_n^k|\times |Q_n^k|} = ((s_{\beta_1\beta_2}^{\alpha_1\alpha_2})_{|P_{\alpha_1}|\times |P_{\alpha_2}|})_{|Q_n^k|\times |Q_n^k|}
\end{equation}
- where
\begin{equation}
\label{e:bab}
s_{\beta_1\beta_2}^{\alpha_1\alpha_2} =  \bigwedge_{(i,\mu),(j,\nu)\in\{(i_1^{\alpha_1},\mu_1^{\beta_1}),\ldots,(i_k^{\alpha_1},\mu_k^{\beta_1})\}\cup\{(i_1^{\alpha_2},\mu_1^{\beta_2}),\ldots,(i_k^{\alpha_2},\mu_k^{\beta_2})\}} b_{\mu\nu}^{ij}
\end{equation}
- where $b_{\mu\nu}^{ij}$ are the components of the given compatibility matrix \ref{e:1ten}. And with the replacement of system \ref{e:mi1} by the following system:
\begin{equation}
\label{e:mi}
\left\{ \begin{array}{rrcl}
1)& \chi_{\beta_1\beta_2}^{\alpha_1\alpha_2} & \Rightarrow & s_{\beta_1\beta_2}^{\alpha_1\alpha_2}  \\
2)& \chi_{\beta_1\beta_2}^{\alpha_1\alpha_2}  & \Rightarrow & \bigvee_{\beta\in[|P_\alpha|]} \chi_{\beta_1\beta}^{\alpha_1\alpha},~\alpha \neq \alpha_2 \\
  &\chi_{\beta_1\beta_2}^{\alpha_1\alpha_2}  & \Rightarrow & \bigvee_{\beta\in[|P_\alpha|]} \chi_{\beta\beta_2}^{\alpha \alpha_2}, ~\alpha\neq \alpha_1 \\ 
3)&true & \Rightarrow & \bigvee_{\beta\in[|P_{\alpha_1}|]} \chi_{\beta\beta}^{\alpha_1\alpha_1} \\ 
\end{array}\right.
\end{equation}
It will even reduce uniform threshold \ref{e:ut1}: everything above can be edited for compatibility matrix \ref{e:sk} and system \ref{e:mi} in assumption $k\geq \kappa_0$, where
\begin{equation}
\label{e:ut0}
\kappa_0 = \min\{n, m\}
\end{equation}
Still, matrix \ref{e:rk} and its derivatives seem to be easier for the preprocessing. 

\section{Compatibility boxes}
\label{s:boxes}

Solution grids in compatibility matrix \ref{e:1ten} are in one-to-one relation with the true assignments satisfying the following Boolean formula\footnote{"$\oplus$" is the XOR operator.}:
\begin{equation}
\label{e:g-1}
g_1 = \bigwedge_{i\in[n]}~\bigoplus_{\mu\in[m_i]}~ \chi_\mu^i~\wedge~\bigwedge_{\neg b_{\mu\nu}^{ij}} (\neg \chi_\mu^i \vee \neg \chi_\nu^j)
\end{equation}
The one-to-one relation is as follows: 
\[
\chi_\mu^i = (\phi(i) = \mu),~i\in[n],\mu\in[m_i]
\]
- where $\phi$ is index function of a solution grid in compatibility matrix \ref{e:1ten}. The $O(m^2n)$-time replacement of all XOR operators in formula $g_1$ by their CNF-expressions\footnote{$x\oplus y\oplus\ldots \oplus z \Leftrightarrow (x\vee y\vee\ldots\vee z)\wedge(\bar{x}\vee\bar{y})\wedge\ldots(\bar{x}\vee\bar{z})\wedge\ldots\wedge(\bar{y}\vee\bar{z})\wedge\ldots$} will one-one reduce formula $g_1$ to CF
\begin{equation}
\label{e:g-2}
g_2 = \bigwedge_{i\in[n]}~\bigvee_{\mu\in[m_i]}~ \chi_\mu^i~\wedge~\bigwedge_{\neg b_{\mu\nu}^{ij}} (\neg \chi_\mu^i \vee \neg \chi_\nu^j)
\end{equation}
And Cook-Karp equivalence transformations\footnote{$x\vee y \vee z \vee \ldots \Leftrightarrow (x\vee y\vee \neg\xi)\wedge(\xi\vee z\vee \ldots)\Leftrightarrow \ldots $} \cite{09,10} will one-one reduce CF $g_2$ to a 3CF. In the 3CF, every couple of the slack literals $\xi$ and $\neg\xi$ can be replaced by two variables $\zeta_1 = \xi$, $\zeta_2 = \neg\xi$, and constrain
\[
\zeta_1 \oplus \zeta_2 = (\zeta_1 \vee \zeta_2)\wedge(\neg\zeta_1 \vee \neg\zeta_2) = true
\]
Because of the XOR operators in $g_1$, these constrains can be reduced:
\[
(\zeta_1 \oplus \zeta_2 = true)|_{g_2} ~\Leftrightarrow ~(\neg\zeta_1 \vee \neg\zeta_2 = true) 
\]
So, substitution $\zeta_1 = \xi$, $\zeta_2 = \neg\xi$ and adding of clauses $\neg\zeta_1 \vee \neg\zeta_2$ will one-one reduce that 3CF into formula \ref{e:g-2} for a compatibility matrix with the compatibility boxes of size $3\times 3$ or less. When needed, the compatibility boxes whose size is less than $3\times 3$ can be always padded\footnote{For example, $x \Leftrightarrow(x\vee \neg \xi)\wedge (x\vee\xi)\Leftrightarrow\ldots$}. Therefore, the following analog of Cook's 3SAT theorem \cite{09} holds:
\begin{theorem}
\label{t:c3sat}
Compatibility matrix \ref{e:1ten} is $O(m^2n^2)$-time one-one reducible to a $O(mn\times mn)$ compatibility matrix with the $3\times 3$ compatibility boxes.
\end{theorem}
The $O(m^2n^2)$-time one-one reduction of the given compatibility matrix \ref{e:1ten} to a compatibility matrix with the $3\times 3$ compatibility boxes may be included in the preprocessing phase. Then, the tabulation/wiring of enumerations \ref{e:Q}, \ref{e:q}, and \ref{e:P} for the $3\times 3$ compatibility boxes will reduce the time-complexity of the compatibility matrix method to $O(m^4n^4)$ and the space-complexity to $O(m^8n^8)$: the $O(m^4n^4)$-depth/width circuit \ref{e:mi13} performs parallel computing of the $O(m^4n^4)$ threads.
\newline\indent
Any particular protocol for enumerations \ref{e:Q}, \ref{e:q}, and \ref{e:P} for compatibility matrices with the $3\times 3$ compatibility boxes will make the sets of the appropriate systems \ref{e:mi1} and \ref{e:mi} (and the sets of their derivatives) the sparse sets. And computation of the implicants (the right sides) in the first groups of implications in the systems will be Karp reduction of the problems allowing the compatibility-matrix encoding to those sparse sets. 

\section{Symmetries}

Compatibility-matrix encoding allows another interpretation which may well serve as the encoding's recipe: there is a \emph{placeholder} for the solution of the given problem; the placeholder is partitioned in $n$ parts which may overlap; there are options for each of the parts - set $O_i$ for part $i$, $m_i = |O_i|$; the options are bound by the $n^2$ 2-ary relations $\rho_{ij}\subseteq O_i\times O_j$ - the diagonal relations $\rho_{ii}$ are the inclusion relations, and the off-diagonal relations $\rho_{ij}$, $i\neq j$, are the \emph{compatibility relations}; the parts and their options are enumerated; and compatibility boxes $B_{ij}$ are the appropriate graphics of relations $\rho_{ij}$. Then, the problem itself is a system of relations:
\begin{equation}
\label{e:ro}
(x_i,x_j)\in \rho_{ij}, ~i,j\in[n] 
\end{equation}
- where variables $x_i\in O_i$ are the unknown options for the placeholder's parts. And, we see, the problems allowing the compatibility-matrix encoding are the only problems which allow binary decomposition \ref{e:ro}.
\begin{xca}
Along with compatibility relations we could use the \emph{contradiction relations}. Basically, the \emph{contradiction matrix} is negation of the compatibility matrix:
\[
\neg B = (\neg B_{ij})_{n\times n} = ((\neg b_{\mu\nu}^{ij})_{m_i\times m_j})_{n\times n}
\]
The contradiction and compatibility matrices are De Morgan dual.
\end{xca}
Relabeling of the placeholder' parts and the parts' options, remodeling of the placeholder and its partition, and reducibility among the problems allowing the compatibility matrices all that may be seen as the symmetries of the compatibility-matrix encoding: the relabeling transforms the given compatibility matrix into a congruent compatibility matrix\footnote{Two square box matrices are congruent if they can be transformed one into other with the permutations of the same rows and columns of boxes and with the permutations of the same rows and columns of components inside of the rows and columns of boxes.}; the remodeling is covered to some extend by Theorem \ref{t:c3sat}; and partition of the problems on consistent and inconsistent is covered to some extend by Theorems \ref{t:c3sat} and \ref{t:mi1} - computational complexity of the partition projection is polynomial in the compatibility matrices' size. Let's see finer partitions.   
\newline\indent
Theorem \ref{t:mi1} gives the general solution of compatibility matrix \ref{e:1ten}. Then, the following Boolean equation gives an exterior rectangular estimation of the relation created by the problem encoded by compatibility matrix \ref{e:1ten}:
\begin{equation}
\label{e:rectangle}
\bigwedge_{i\in [n]} ~\bigvee_{\theta_{\mu\mu}^{ii}} \chi_\mu^i = true
\end{equation}
And this estimation is exact: Boolean equation \ref{e:rectangle} is consistent iff the encoded problem is consistent. The vertices of rectangle \ref{e:rectangle} are solutions of the following Boolean equation:
\begin{equation}
\label{e:vertices}
\bigwedge_{i\in [n]} ~\bigoplus_{\theta_{\mu\mu}^{ii}} \chi_\mu^i = true
\end{equation}
This equation is the affine case in Schafer's classification \cite{11}. And its solutions can be expressed either by system of linear inequalities or even by system of linear equations in Euclidean space {\bf R}$^{\sum_{i\in[n]}m_i}$ = {\bf R}$^{O(mn)}$:
\begin{equation}
\label{e:linear}
\left\{\begin{array}{rl}
\sum_{\mu\in [m_i]} x_\mu^i = 1, & i\in[n] \\
x_\mu^i = 0, & \neg \theta_{\mu\mu}^{ii} \\
0 \leq x_\mu^i \leq 1 & \\
\end{array}\right.
~\Leftrightarrow~
\left\{\begin{array}{rl}
\sum_{\mu\in [m_i]} x_\mu^i = 1, & i\in[n] \\
x_\mu^i = 0, & \neg \theta_{\mu\mu}^{ii} \\
& \\
\end{array}\right.
\end{equation}
Relabeling of the placeholder's parts and the parts' options in system \ref{e:ro} will rotate coordinates in Euclidean space {\bf R}$^{\sum_{i\in[n]}m_i}$. And sets of solutions of these linear systems will "rotate" after the coordinates. Because these solution sets are asymmetric in general, systems \ref{e:linear} are asymmetric in the sense of Yannakakis' theorem \cite{12}. 
\newline\indent
With Theorem \ref{t:c3sat}, system of linear equations \ref{e:linear} in the appropriate Euclidean space {\bf R}$^{O(mn)}$ can be obtained in the time polynomial in $mn$. The set of all such linear systems can be partitioned with the congruence relation - projection of this partition is linear in $mn$. And each of these equivalence classes can be presented by such a system in which
\[
3 ~\geq~ \sum_{\theta_{\mu\mu}^{11}} 1 ~\geq ~\sum_{\theta_{\mu\mu}^{22}} 1 ~\geq~ \ldots ~\geq~ \sum_{\theta_{\mu\mu}^{O(mn)O(mn)}} 1 ~\geq ~ 0
\]  
Therefore, the equivalence classes in the partition of the problems allowing the compatibility-matrix encoding based on the exterior estimation \ref{e:linear} and the relabeling of the placeholder's parts and the parts' options in system \ref{e:ro} are another sparse set along with the sparse sets of Boolean systems \ref{e:mi1}, \ref{e:mi}, and their derivatives.
\newline\indent
Let's notice, the sparse sets described in Section \ref{s:boxes} consist of the "universal algorithms" while the sparse set described in this section consists of the "encodings". In other words, the hardness of the problems allowing the compatibility matrix encoding can be dissolved by the "lucky choice" of the placeholder, the placeholder's partition, and the partition's labeling.
\begin{xca}
Boolean equation \ref{e:vertices} is Karp reducible to 2SAT through the replacement of the XOR operators by their CNF-expression and usual self-reducibility of the resulting CNF. Therefore, for the problems allowing the compatibility-matrix encoding, all cases in Schafer's classification are Karp reducible each to other except the case reserved for the intractable problems. But, we saw, the intractable problems lay outside of the compatibility matrix method's domain.
\end{xca}
We have discussed the symmetry based on the exact exterior rectangular estimations \ref{e:rectangle} and \ref{e:vertices}. Yet, there are the similar interior estimations.
\newline\indent
The solution grids in compatibility matrix \ref{e:1ten} are in one-to-one relation with the true assignments satisfying Boolean formula \ref{e:g-1}. And formula \ref{e:g-1} is self-reducible with the following equivalence transformations: 
\[
(x\oplus y\oplus\ldots\oplus z)\wedge(\bar{x}\vee X)\wedge(\bar{y}\vee Y)\wedge\ldots\wedge(\bar{z}\vee Z)~\Leftrightarrow~X\vee Y \vee\ldots\vee Z
\]
- where variables $x,y,\ldots,z$ are independent and formulae $X,Y,\ldots,Z$ are independent of variables $x,y,\ldots,z$. These equivalence transformations will transform formula \ref{e:g-1} into the following formula (computational complexity of this direct reduction will be exponential-time, in general):
\begin{equation}
\label{e:g-11}
g_1\acute{} = \bigwedge_{i\in N \subseteq [n]}~ (\bigoplus_{\mu\in[m_i]}\chi_\mu^i~\wedge~\bigwedge_{\mu\in M_i\subseteq [m_i]} \neg\chi_\mu^i) 
\end{equation}
- where $N\neq\emptyset$ and some $M_i = [m_i]$ in the case of unsatisfiable formula \ref{e:g-1}.
\begin{xca}
\label{x:2222}
Formula \ref{e:g-1} for the compatibility matrix from Exercise \ref{x:222}:
\[
\begin{array}{rl}
g_1 = & (\chi_1^1 \oplus \chi_2^1)\wedge(\bar{\chi}_1^1\vee \bar{\chi}_2^2)\wedge(\bar{\chi}_1^1\vee\bar{\chi}_1^3)\wedge(\bar{\chi}_2^1\vee\bar{\chi}_1^2)\wedge(\bar{\chi}_2^1\vee\bar{\chi}_2^3) \\
\wedge & (\chi_1^2\oplus\chi_2^2)\wedge(\bar{\chi}_1^2\vee\bar{\chi}_2^3)\wedge(\bar{\chi}_2^2\vee\bar{\chi}_1^3) \\
\wedge & (\chi_1^3\oplus\chi_2^3) \\
\Leftrightarrow & (\bar{\chi}_2^2\wedge\bar{\chi}_1^3\vee \bar{\chi}_1^2\wedge\bar{\chi}_2^3) \\
\wedge & (\chi_1^2\oplus\chi_2^2)\wedge(\bar{\chi}_1^2\vee\bar{\chi}_2^3)\wedge(\bar{\chi}_2^2\vee\bar{\chi}_1^3) \\
\wedge & (\chi_1^3\oplus\chi_2^3) \\
\wedge & (\chi_1^3\oplus\chi_2^3) \\
\Leftrightarrow & (\bar{\chi}_2^2\vee\bar{\chi}_1^2)\wedge(\bar{\chi}_1^3\vee\bar{\chi}_1^2) \wedge(\bar{\chi}_2^2\vee\bar{\chi}_2^3)\wedge(\bar{\chi}_1^3\vee\bar{\chi}_2^3) \\
\wedge & (\chi_1^2\oplus\chi_2^2)\wedge(\bar{\chi}_1^2\vee\bar{\chi}_2^3)\wedge(\bar{\chi}_2^2\vee\bar{\chi}_1^3) \\
\wedge & (\chi_1^3\oplus\chi_2^3) \\
\hline
\Leftrightarrow &  (\chi_1^2\oplus\chi_2^2)\wedge(\bar{\chi}_1^2\vee\bar{\chi}_2^3)\wedge(\bar{\chi}_2^2\vee\bar{\chi}_1^3)\wedge(\bar{\chi}_1^3\vee\bar{\chi}_1^2) \wedge(\bar{\chi}_2^2\vee\bar{\chi}_2^3) \\
\wedge & (\chi_1^3\oplus\chi_2^3) \\
\Leftrightarrow &  (\bar{\chi}_2^3\wedge\bar{\chi}_1^3\vee\bar{\chi}_1^3\wedge\bar{\chi}_2^3) \\
\wedge & (\chi_1^3\oplus\chi_2^3) \\
\hline
\Leftrightarrow & (\chi_1^3\oplus\chi_2^3)\wedge \bar{\chi}_1^3\wedge\bar{\chi}_2^3 ~ = g_1\acute{}
\end{array}
\]
We see, $M_3 = [m_3] = \{1,2\}$ in formula \ref{e:g-11} for this compatibility matrix. In the case of a compatibility matrix with the larger compatibility boxes, the length of the newly created all-negative OR-clauses can increase at the beginning of the equivalence transformations. But, the length is bounded by the number of remaining XOR-clauses. So, it will decrease at the end.
\end{xca}
Boolean equation $g_1\acute{} = true$ defines vertices of rectangle
\[
\bigwedge_{i\in N} ~ \bigvee_{\mu\in[m_i]-M_i} \chi_\mu^i = true
\]
This dimensionally degenerated rectangle lays inside of the solution set of system \ref{e:ro}. And this interior estimation is exact: there are solution grids in compatibility matrix \ref{e:1ten} iff this rectangle is not empty. Thus, all above for exterior rectangle \ref{e:rectangle} holds for this interior rectangle.
\begin{xca}
\label{x:aonot}
Formula \ref{e:g-11} is not unique and depends on which XOR-clauses we choose to exclude from formula \ref{e:g-1}. Suppose, we have excluded the $i$th XOR-clause. Then,
\begin{equation}
\label{e:aonot}
\neg \chi_\mu^i~\Leftarrow~f_\mu^i,~\mu\in[m_i]
\end{equation}
- where each $f_\mu^i$ is an exponential-size all-positive DNF on the remaining variables. In full agreement with Rossman's theorem \cite{28}, these implications create an AND-OR-NOT circuit of the linear depth and width, en masse. Unlike circuits \ref{e:mi13}, this family of circuits is nonuniform.
\newline\indent 
Implications \ref{e:aonot} are an implicit function defined by system \ref{e:ro}. These intrinsic dependencies will cause clustering and phase transitions in the numerical experiments with the random search, and they can be identified with the statistical methods due to equality PSPACE = IP.  
\end{xca}

\section{Similarity to quantum computer}

System \ref{e:ro} reassembles quantum computer: variables $x_i$ act like qubits, compatibility relations $\rho_{ij}$ act like entanglement, and the system itself acts like superposition. And the following exact deterministic algorithm, which is called \emph{depletion by multiplication}, reassembles Grover's algorithm \cite{13}.
\begin{theorem}
\label{t:multi}
For compatibility matrix \ref{e:1ten}:
\newline
1). Let's compute compatibility matrix \ref{e:sk} - matrix $S(k)$;
\newline
2). Let's iterate compatibility boxes in compatibility matrix $S(k)$ and recompute the current box with formula\footnote{We use the following Boolean matrix multiplication "$\cdot$": 
\[
(a_{i\mu})\cdot(b_{\mu j}) = (\bigvee_\mu a_{i\mu}\wedge b_{\mu j})
\]
- where the number of columns in $(a_{i\mu})$ equals the number of rows in $(b_{\mu j})$. And formula \ref{e:multi} is formula of the power of box matrix $S(k)$ in the sense of the mixed Boolean box matrix multiplication: the components are treated in the existential sense while the boxes are treated in the universal sense. "$\cdot$" has  higher precedence than "$\wedge$". }
\begin{equation}
\label{e:multi}
S_{\alpha_1\alpha_2}(k) = \bigwedge_{\alpha=1}^{|Q_n^k|} S_{\alpha_1\alpha}(k)\cdot S_{\alpha\alpha_2}(k)
\end{equation}
3). Let's loop these iterations until compatibility matrix $S(k)$ got finalized.
\newline
The final value of compatibility matrix $S(k)$ is general solution of compatibility matrix \ref{e:sk} if $k\geq 0.5 \kappa_0$, where $\kappa_0$ is uniform threshold \ref{e:ut0}. 
\end{theorem}
\begin{proof}
Conjunction \ref{e:multi} depletes compatibility matrix $S(k)$ because the diagonal compatibility boxes are diagonal Boolean matrices. This algorithm will stop after $O((m^k\mbox{C}_n^k)^2)$ loops at most because there are $(m^k\mbox{C}_n^k)^2$ $true$-components in compatibility matrix $S(k)$ at most. 
\newline\indent
Due to the compatibility matrices' symmetry,
\[
\bigwedge_{\alpha=1}^{|Q_n^k|} S_{\alpha_1\alpha}(k)\cdot S_{\alpha\alpha_2}(k) = \bigwedge_{\alpha=1}^{|Q_n^k|} (\bigvee_{\beta\in |P_\alpha|}s_{\beta_1\beta}^{\alpha_1\alpha}\wedge s_{\beta_2\beta}^{\alpha_2\alpha})_{|P_{\alpha_1}|\times |P_{\alpha_2}|} 
\]
Therefore, solution grids in compatibility matrix \ref{e:sk} are invariants under transformation \ref{e:multi}. On other hand, due to equalities \ref{e:bab}, the solution grids in compatibility matrix \ref{e:sk} are in one-to-one relation with the solution grids in compatibility matrix \ref{e:1ten}:
\[
\psi(\alpha) = \beta~\Leftrightarrow~ \bigwedge_{\iota\in[k]} (\phi(i_\iota^\alpha) = \mu_\iota^\beta)
\]
- where $\psi$ and $\phi$ are index functions of the appropriate solution grids in compatibility matrices \ref{e:sk} and \ref{e:1ten}. Then (see Exercise \ref{x:observ}), while compatibility matrix $S(k)$ for $k\geq \kappa_0/2$ contains noise, there always will be such indices $\alpha_1,\alpha_2,\alpha\in [|Q_n^k|]=[\mbox{C}_n^k]$ that $s_{\beta_1\beta_2}^{\alpha_1\alpha_2} = true$ and
\[
\bigvee_{\beta\in |P_\alpha|}s_{\beta_1\beta}^{\alpha_1\alpha}\wedge s_{\beta_2\beta}^{\alpha_2\alpha} ~\Rightarrow ~\bigvee_{\mu\in[m_{i_0}]} b_{\mu\mu}^{i_0i_0}\wedge \bigwedge_{\iota\in[k]} b_{\mu_\iota\mu}^{i_\iota i_0} ~= ~false
\]
for some $i_0\in q_\alpha$, $i_1,i_2,\ldots,i_k\in q_{\alpha_1} \cup q_{\alpha_2}$, and the appropriate indices $\mu_\iota,~\iota\in[k],$ from $k$-tuples $p_{\beta_1}\in P_{\alpha_1}$ and $p_{\beta_2}\in P_{\alpha_2}$. Therefore, when $k\geq \kappa_0/2$, the looping will not stop while there is noise in matrix $S(k)$.
\end{proof}
Because of the box conjunctions in formula \ref{e:multi}, the number of loops in the algorithm cannot be greater than the length of the longest path in the graph (with loops) whose adjacency matrix is compatibility matrix \ref{e:sk} after the replacement of $false$ by $0$ and $true$ by $1$. It is C$_n^k$-partite graph. So, the number of loops in the depletion by multiplication algorithm is $O(\mbox{C}_n^k)$. And the time-complexity of the depletion by multiplication algorithm is $O((m^k\mbox{C}_n^k)^4)$ while its space-complexity is $O((m^k\mbox{C}_n^k)^2)$, $k\geq \kappa_0/2$. 
\newline\indent
The depletion by multiplication is filtering of the compatibility boxes each against others.
Before the filtering, we may use Theorem \ref{t:c3sat} and split the given compatibility matrix \ref{e:1ten} into the $O(m^2n^2)$ compatibility boxes of size $3\times 3$. It will make $\kappa_0 = 3$ and $k\geq 2 > 3/2$. Then, in full compliance with Baker, Gill, and Solovey's theorem \cite{14}, the number of loops in the depletion by multiplication will be controllable with $k$ from $O(m^2n^2)$ when $k=2$, up to $O(2^{mn})$ when $k \approx mn/2$, and down to $O(1)$ when $k\approx mn$. 
\newline\indent
The splitting of the compatibility matrix and selection of $k= 2$ will minimize the time-complexity of the depletion by multiplication algorithm which will be $O(m^{8}n^{8})$ and the space-complexity of the algorithm which will be $O(m^4n^4)$. The time- and space-complexities swapped in the monotone circuit \ref{e:mi12} and the depletion by multiplication "quantum algorithm."  So, it seems, these approaches are on Pareto frontier of the computational complexity of the compatibility matrix method.
\newline\indent
Indeed, the depletion by multiplication works as well for compatibility matrix \ref{e:rk} and $k\geq 0.5 \kappa_1$, where $\kappa_1$ is uniform threshold \ref{e:ut1}.
\begin{xca}
The depletion by multiplication works directly on compatibility matrix \ref{e:1ten} when $m \leq 2$, where $m$ is maximum \ref{e:11max}. For example, for the compatibility matrix from Exercise \ref{x:222},
\[
B_{12} = B_{11}\cdot B_{12}\wedge B_{12}\cdot B_{22}\wedge B_{13}\cdot B_{32} =
\]
\[
=\left(\begin{array}{cc}
1 & 0 \\
0 & 1 \\
\end{array}\right)
\cdot
\left(\begin{array}{cc}
1 & 0 \\
0 & 1 \\
\end{array}\right)
\wedge
\left(\begin{array}{cc}
1 & 0 \\
0 & 1 \\
\end{array}\right)
\cdot
\left(\begin{array}{cc}
1 & 0 \\
0 & 1 \\
\end{array}\right)
\wedge
\left(\begin{array}{cc}
0 & 1 \\
1 & 0 \\
\end{array}\right)
\cdot
\left(\begin{array}{cc}
1 & 0 \\
0 & 1 \\
\end{array}\right) =
\]
\[
=\left(\begin{array}{cc}
1 & 0 \\
0 & 1 \\
\end{array}\right)
\wedge
\left(\begin{array}{cc}
1 & 0 \\
0 & 1 \\
\end{array}\right)
\wedge
\left(\begin{array}{cc}
0 & 1 \\
1 & 0 \\
\end{array}\right)
=
\left(\begin{array}{cc}
0 & 0 \\
0 & 0 \\
\end{array}\right)
\]
There is no need to continue the depletion because it will just propagate this $false$-box $B_{12}$ all over the rest of compatibility matrix. It is a \emph{pattern of unsatisfiability} when formula \ref{e:multi} produces (and it will always do so sooner or later for the inconsistent problems) a compatibility box completely filled with $false$. When the pattern of unsatisfiability is detected, the depletion by multiplication may be stopped with the decision "No."
\end{xca}

\section{Application to the P vs NP problem}
\label{s:x}

\subsection{SAT}
\label{s:sat}

Let $f$ be the given CNF:
\begin{equation}
\label{e:f}
f = \bigwedge_{i\in[n]}~ \bigvee_{\mu\in m_i} L_\mu^i,~L_\mu^i\in\{x_\alpha,\neg x_\alpha~|~\alpha\in A \}
\end{equation}
- where $x_\alpha$ are independent Boolean variables, and $L_\mu^i$ are the variables' literals. The problem is to decide whether or not $f$ is satisfiable. 
\newline\indent
The following box matrix $((b_{\mu\nu}^{ij})_{m_i\times m_i})_{n\times n}$ is a compatibility matrix for SAT instance \ref{e:f}:
\begin{equation}
\label{e:cnf1}
b_{\mu\nu}^{ij} = (i \neq j \vee \mu = \nu) \wedge (i = j \vee L_\mu^i \neq \neg L_\nu^j)
\end{equation}
Solution grids in this compatibility matrix are in one-to-one relation with the index-distinct implicants in the following DF of formula \ref{e:f}:
\[
f = \bigvee_{(\mu_1,\mu_2,\ldots,\mu_n)\in[m_1]\times[m_2]\times\ldots\times [m_n]} L_{\mu_1}^1 \wedge L_{\mu_2}^2 \wedge\ldots\wedge L_{\mu_n}^n
\]
The one-to-one relation is:
\[
\phi ~\stackrel{1-1}{\longleftrightarrow}~L_{\phi(1)}^1 \wedge L_{\phi(2)}^2 \wedge\ldots\wedge L_{\phi(n)}^n
\]
- where $\phi$ is index function of a solution grid.
\newline\indent
For SAT, Theorem \ref{t:c3sat} becomes exactly the Cook's 3SAT theorem. Then, Theorem \ref{t:mi1} positively resolves the  P vs NP problem and more:
\[
\mbox{P = NP }\subseteq \mbox{ P/poly}
\]
And, in full compliance with Fortune and Mahaney's theorems \cite{15,16}, monotone circuits \ref{e:mi13} and the congruence classes of linear systems \ref{e:linear} are examples of the NP-complete sparse set. Also, in full agreement with the Karp-Lipton-Sipser \cite{29} and Kannan's \cite{31} theorems, the size  of the implicit function-circuit \ref{e:aonot} is quadratic: linear depth times linear width.
\begin{xca}
\label{x:2221}
There is not CNF whose compatibility matrix \ref{e:cnf1} equals the compatibility matrix from Exercise \ref{x:222}.
\newline\indent
Really, let's assume the opposite. Let $f$ be a CNF whose compatibility matrix \ref{e:cnf1} is the above matrix $B$. Then,
\[
f = (L_1^1\vee L_2^1) \wedge (L_1^2\vee L_2^2) \wedge (L_1^3\vee L_2^3)
\]
Then, there is the following contradiction:
\[
\left. \begin{array}{rcl}
b_{12}^{12} = false & \Rightarrow & L_1^1 = \neg L_2^2 \\
b_{11}^{13} = false & \Rightarrow & L_1^1 = \neg L_1^3 \\
b_{21}^{23} = false & \Rightarrow & L_2^2 = \neg L_1^3 \\
\end{array}\right\} ~ \Rightarrow ~ \neg L_2^2 = L_1^1 = \neg L_1^3 = L_2^2
\]
Still, there is CNF \ref{e:g-2} for this compatibility matrix (see Exercise \ref{x:2222}).
\end{xca}

\subsection{Subgraph isomorphism}
\label{s:1si}
Let $g_1=(V_1,E_1)$ and $g_2=(V_2,E_2)$ be two graphs. The problems is to decide whether or not there is injection $\psi: V_1\rightarrow V_2$ such that
\[
(u_1,u_2)\in E_1~\Rightarrow~(\psi(u_1),\psi(u_2))\in E_2
\]
\indent
Let's arbitrarily enumerate the graphs' vertex sets:
\[
V_1 = \{u_1,u_2,\ldots,u_{|V_1|}\},~V_2=\{w_1,w_2,\ldots,w_{|V_2|}\}
\]
Let $A_1=(a_{ij}^1)_{|V_1|\times |V_1|}$ and $A_2=(a_{ij}^2)_{|V_2|\times |V_2|}$ be the appropriate \emph{Boolean adjacency matrices} of graphs $g_1$ and $g_2$:
\[
a_{ij}^1 = ((u_i,u_j)\in E_1),~ a_{ij}^2 = ((w_i,w_j)\in E_2)
\]
The following box matrix $B=(B_{ij})_{|V_1|\times |V_1|}$ is a compatibility matrix for the Subgraph isomorphism problem:
\begin{equation}
\label{e:1si}
B_{ij}=\left\{\begin{array}{rl}
I_{|V_2|},& i = j \\
\neg I_{|V_2|}, & i\neq j ~\wedge~\neg a_{ij}^1 \\
A_2, & i \neq j~\wedge~a_{ij}^1 \\
\end{array}\right.
\end{equation}
-  where $I_{|V_2|}$ is the $|V_2|\times |V_2|$ Boolean identity matrix\footnote{All diagonal components equal $true$, and all off-diagonal components equal $false$.}, and matrix $\neg I_{|V_2|}$ is the per-component negation of matrix $I_{|V_2|}$. Solution grids in this compatibility matrix are in one-to-one relation with the injections in question:
\[
\phi~\stackrel{1-1}{\longleftrightarrow}~\psi = (u_1\mapsto w_{\phi(1)},u_2\mapsto w_{\phi(2)},\ldots,u_{|V_1|}\mapsto w_{\phi(|V_1|)})
\]
- where $\phi$ is index function of a solution grid.
\begin{xca}[Clique]
\label{x:clique}
For the $k$-Clique problem adjacency matrix of graph $g_1$ is
\[
A_1 = \left ( \begin{array}{ccccccc}
0 & 1 & 1 & \ddots & 1 & 1 & 1 \\
1 & 0 & 1 & \ddots & 1 & 1 & 1 \\
1 & 1 & 0 & \ddots & 1 & 1 & 1 \\
\ddots & \ddots & \ddots & \ddots & \ddots & \ddots & \ddots \\
1 & 1 & 1 & \ddots & 0 & 1 & 1 \\
1 & 1 & 1 & \ddots & 1 & 0 & 1 \\
1 & 1 & 1 & \ddots & 1 & 1 & 0 \\
\end{array}\right )_{k\times k}
\]
And compatibility matrix \ref{e:1si} is
\[
B = \left ( \begin{array}{ccccccc}
I_{|V_2|} & A_2 & A_2 & \ddots & A_2 & A_2 & A_2 \\
A_2 & I_{|V_2|} & A_2 & \ddots & A_2 & A_2 & A_2 \\
A_2 & A_2 & I_{|V_2|} & \ddots & A_2 & A_2 & A_2 \\
\ddots & \ddots & \ddots & \ddots & \ddots & \ddots & \ddots \\
A_2 & A_2 & A_2 & \ddots & I_{|V_2|} & A_2 & A_2 \\
A_2 & A_2 & A_2 & \ddots & A_2 & I_{|V_2|} & A_2 \\
A_2 & A_2 & A_2 & \ddots & A_2 & A_2 & I_{|V_2|} \\
\end{array}\right )_{k\times k}
\]
This compatibility-matrix encoding is Karp reduction of the $k$-Clique problem to the $k$-Clique instance in the $k$-partite graph whose adjacency matrix is matrix $B$ above (we just nil the diagonal components). That is a special case. And its polynomial circuit complexity completely agrees with Razborov's super-polynomial estimations \cite{17} (later tightened by Alon and Boppana \cite{18}) of the circuit complexity of the worst case of Clique.
\newline\indent
The solution grids existence problem itself is an instance of the $n$-Clique problem in a $n$-partite graph with $O(mn)$ vertices: we just nil in compatibility matrix \ref{e:1ten} all its diagonal components to obtain an adjacency matrix of that graph. The $n$-partiteness of this model is the major benefit of the compatibility-matrix encoding (See Exercise \ref{x:observ}).
\end{xca}
\begin{xca}[Graph isomorphism]
Graph isomorphism problem is the Subgraph isomorphism instance when $|V_1| = |V_2|$ and $|E_1| = |E_2|$.
\end{xca}
\begin{xca}[Path and Cycle]
\label{x:pc}
The $k$-Path/Cycle problem is a Subgraph isomorphism instance when graph $g_1$ has adjacency matrix
\[
A_1 = \left ( \begin{array}{ccccccc}
0 & 1 & 0 & \ddots & 0 & 0 & c \\
1 & 0 & 1 & \ddots & 0 & 0 & 0 \\
0 & 1 & 0 & \ddots & 0 & 0 & 0 \\
\ddots & \ddots & \ddots & \ddots & \ddots & \ddots & \ddots \\
0 & 0 & 0 & \ddots & 0 & 1 & 0 \\
0 & 0 & 0 & \ddots & 1 & 0 & 1 \\
c & 0 & 0 & \ddots & 0 & 1 & 0 \\
\end{array}\right )_{k\times k}
\]
- where $c=0$ for the Path and $c=1$ for the Cycle. Compatibility matrix \ref{e:1si} is 
\[
B = \left ( \begin{array}{ccccccc}
I_{|V_2|} & A_2 & \neg I_{|V_2|} & \ddots & \neg I_{|V_2|} & \neg I_{|V_2|} & C \\
A_2 & I_{|V_2|} & A_2 & \ddots & \neg I_{|V_2|} & \neg I_{|V_2|} & \neg I_{|V_2|} \\
\neg I_{|V_2|} & A_2 & I_{|V_2|} & \ddots & \neg I_{|V_2|} & \neg I_{|V_2|} & \neg I_{|V_2|} \\
\ddots & \ddots & \ddots & \ddots & \ddots & \ddots & \ddots \\
\neg I_{|V_2|} & \neg I_{|V_2|} & \neg I_{|V_2|} & \ddots & I_{|V_2|} & A_2 & \neg I_{|V_2|} \\
\neg I_{|V_2|} & \neg I_{|V_2|} & \neg I_{|V_2|} & \ddots & A_2 & I_{|V_2|} & A_2 \\
C & \neg I_{|V_2|} & \neg I_{|V_2|} & \ddots & \neg I_{|V_2|} & A_2 & I_{|V_2|} \\
\end{array}\right )_{k\times k}
\]
- where $C= \neg I_{|V_2|}$ for the Path and $C=A_2$ for the Cycle.
\newline\indent
The Hamiltonian path/cycle problem is the instance when $k = |V_2|$.
\end{xca}

\subsection{QSAT}

Let $F$ be the following quantified CNF in prenex normal form:
\begin{equation}
\label{e:F}
F = Q_1 x_{1} Q_2 x_{2} \ldots Q_k x_{k} f
\end{equation}
- where  matrix $f$ is CNF \ref{e:f}; $Q_1, Q_2, \ldots, Q_k\in\{\forall,\exists\}$ are quantifiers; the number of quantifiers equal the number of variables in formula $f$, $k=|A|$; and variables $x_1, x_2,\ldots, x_{k}$ in formula $f$ are enumerated in the order of the quantification. The problem is to compute $F$.
\newline\indent
The following CF is an equivalent of CNF $f$:
\[
f\acute{~} = f ~\wedge~\bigwedge_{l=1}^k (x_l\vee\bar{x}_l) ~\Leftrightarrow~f
\]
Let's compute truth table for each clause $c_i$ in CF $f\acute{~}$, $i\in[n+k]$:
\[
T_i = \begin{array}{|c|c|c|}
\hline
\mbox{\#} & \mbox{True assignments} & \mbox{Clause $c_i$} \\
\hline
\hline
1 & \mbox{The $1$st true assignment to} & \mbox{The $1$st meaning} \\
& \mbox{the variables in clause $c_i$} & \mbox{ of clause $c_i$} \\
\hline
\vdots & \vdots & \vdots \\
\hline
|T_i| & \mbox{The $|T_i|$th true assignment to} & \mbox{The $|T_i|$th meaning} \\
& \mbox{the variables in clause $c_i$} & \mbox{ of clause $c_i$} \\
\hline
\end{array}
\] 
- where length  $|T_i|$ of table $T_i$ equals $2^{|c_i|}$ for the clause $c_i$ in CNF $f$ ($|c_i|$ is the length of clause $c_i$ and $i\in[n]$), and $|T_i| = 2$ for the added clauses $x_l\vee\bar{x}_l$ ($l\in[k],~i=n+1,n+2,\ldots,n+k$). In the terms of system \ref{e:ro}, the numbers of strings in table $T_i$ will be options for the parts in the solution partition aka clauses $c_i$. And the appropriate compatibility relations are: the $\mu$th string in truth table $T_i$ and the $\nu$th string in truth table $T_j$ are compatible if meanings of all variables in these true assignments do not contradict each other (the same variables have the same true assignment), and the $\mu$th and $\nu$the meanings of clauses $c_i$ and $c_j$ both are $true$. Let's graph all these compatibility relations in the appropriate spaces $[|T_i|]\times [|T_j|]$ and aggregate those graphics in a box matrix in accordance with the clauses' indices:
\begin{equation}
\label{e:bbb}
B = (B_{ij})_{n\times n} = ((b_{\mu\nu}^{ij})_{|T_i|\times |T_i|})_{(n+k)\times (n+k)}
\end{equation}
Compatibility boxes in compatibility matrix $B$ are arranged as follows:
\[
B = \left(~\begin{array}{c|c}
\mbox{Compatibility boxes} & \mbox{Mixed} \\
\mbox{for clauses in CNF $f$} & \mbox{compatibility}\\
\mbox{in the clauses' order} & \mbox{boxes} \\
\hline
\mbox{Mixed} & \mbox{Compatibility boxes} \\
\mbox{compatibility} & \mbox{for clauses $x_l\vee\bar{x}_l$,} \\
\mbox{boxes} & l=1,2,\ldots, k \\
\end{array}~\right)_{(n+k)\times (n+k)}
\]
Solution grids in compatibility matrix $B$ are in one-to-one relation with the true assignments satisfying both formulae $f\acute{~}$ and $f$:
\[
\phi~\stackrel{1-1}{\longleftrightarrow}~ \left\{\begin{array}{l}
\mbox{Variables in clause $c_i$ have} \\
\mbox{their $\phi(i)$th true assignment from} \\
\mbox{the truth table for clause $c_i$} \\
\end{array}\right.
\]
- where $\phi$ is index function of a solution grid in compatibility matrix $B$ and, thus, the same variables in different clauses have the same assignments.
\newline\indent
QCNF \ref{e:F} can be expressed by SAT instance \ref{e:g-2} for compatibility matrix \ref{e:bbb} subject to the following constrains on compatibility boxes $B_{ij}, ~i,j>n$ (these $2\times 2$ compatibility boxes are located in the lower-right corner, and these constrains express leaves in the splitting tree of formula $F$):
\[
\begin{array}{|c||c|c|}
\hline
i < j & Q_{j-n} = \forall & Q_{j-n} = \exists \\
\hline\hline
Q_{i-n} = \forall & b_{11}^{ij}\wedge b_{12}^{ij}\wedge b_{21}^{ij}\wedge b_{22}^{ij} = true & (b_{11}^{ij}\vee b_{12}^{ij})\wedge (b_{21}^{ij}\vee b_{22}^{ij}) = true \\
\hline
Q_{i-n} = \exists & b_{11}^{ij}\wedge b_{12}^{ij}\vee b_{21}^{ij}\wedge b_{22}^{ij} = true & b_{11}^{ij}\vee b_{12}^{ij}\vee b_{21}^{ij}\vee b_{22}^{ij} = true  \\
\hline
\end{array}
\]
The following substitutions may be done in these constrains:
\[
b_{\mu\nu}^{ij} = \chi_\mu^i\wedge\chi_\nu^j
\]
- where $\chi_\mu^i$ and $\chi_\nu^j$ are the appropriate variables in formula \ref{e:g-2} for compatibility matrix \ref{e:bbb}. And the constrains may be replaced by their CNF-expressions and added to formula \ref{e:g-2}. There will be $O(m^2n^2)$ clauses in the resulting CNF, where $O(mn)$ is the size of matrix $f$. And satisfiability of that CNF will indicate whether or not $F=true$. Therefore,
\[
\mbox{PSPACE = NP}
\] 
- we have Karp reduced QSAT to SAT. 
\begin{xca}
Formulae \ref{e:multi}, \ref{e:mi1}, and their derivatives perform parallel testing of all guesses. It makes the compatibility matrix method similar to NDTM. And alternation of operators "$\vee$" and "$\wedge$" in the formulae makes the method similar to Chandra and Stockmeyer's ATM \cite{26,27}.
\end{xca}

\subsection{Reachability}
\label{s:wpc}

Let $g=(V,E)$ be a (multi)digraph (loops are allowed). The problem is to decide whether or not there are $k$-walks and $k$-paths starting in $V_1\subseteq V$ and finishing in $V_2\subseteq V$.
\newline\indent
Let $A = (a_{\mu\nu})_{|V|\times |V|}$ be a Boolean adjacency matrix of digraph $g$ (see Subsection \ref{s:1si}). Let's define powers\footnote{We use the "$\cdot$" Boolean matrix multiplication: $(a_{i\mu})\cdot (b_{\mu j}) = (\bigvee_\mu a_{i\mu}\wedge b_{\mu j})$.} of matrix $A$:
\[
A^0 = I_{|V|};~ A^p = A^{p-1}A,~p>0;~A^p = (A^{-p})^T,~p < 0
\]
- where $I_{|V|}$ is the $|V|\times |V|$ Boolean identity matrix. Let's $D_1$ and $D_2$ be Boolean diagonal matrices indicating sets $V_1$ and $V_2$ appropriately. Then, the following box matrix is a compatibility matrix for the problem:
\[
B = (B_{ij})_{(k+1)\times (k+1)}
\]
- where 
\begin{equation}
\label{e:w}
B_{ij} = \left\{\begin{array}{rl}
D_1, & i = j = 1 \\
I_{|V|}, & i = j \notin \{1,k+1\} \\
D_2, & i = j = k+1 \\
A^{j-i},& i\neq j \mbox{ and it is the Walk problem} \\
\neg I_{|V|}\wedge A^{j-i}, & i \neq j \mbox{ and it is the Path problem} \\
\end{array}\right.
\end{equation}
Solution grids in this compatibility matrix are in one-to-one relation with the $k$-walks/paths in digraph $g$:
\begin{equation}
\label{e:one-to-one}
\phi~\stackrel{1-1}{\longleftrightarrow}~(v_{\phi(1)},v_{\phi(2)},\ldots,v_{\phi(k)},v_{\phi(k+1)})
\end{equation}
- where $\phi$ is index function of a solution grid. This  example will help us to find domain of the compatibility matrix method and more.
\newline\indent
Let $M=(\Sigma,\Gamma,Q,\delta,Q_{\mbox{\scriptsize init}},Q{\mbox{\scriptsize accept}},Q{\mbox{\scriptsize reject}})$ be a Turing machine, deterministic or not:
\[
\delta \subseteq (Q - (Q{\mbox{\scriptsize accept}}\cup Q{\mbox{\scriptsize reject}}))\times \Gamma \times Q \times \Gamma \times \{\leftarrow,\rightarrow\}
\]
Let function $t_M: s\in\Sigma^* \mapsto t_M(s)\in \mbox{\bf N}\cup\{\infty\} = \{0,1,2,\ldots,\infty\}$ be the time during which machine $M$ computes strings $s\in\Sigma^*$ and accept or reject them. Let $T_M$ be the time-complexity of machine $M$:
\[
T_M = \sup \{ t_M(s)~|~s\in\Sigma^*\}
\]
Let's assume that machine $M$ is such that
\[
T_M < \infty
\] 
Then, we can compute compatibility matrix \ref{e:w} of the $T_M$-walks for the following digraph $g$ and sets $V_1$ and $V_2$:
\[
g = (Q, Q\times Q),~V_1 = Q_{\mbox{\scriptsize init}}, ~V_2 = Q{\mbox{\scriptsize accept}}
\]
Let $B$ be that compatibility matrix:
\[
B = ((b_{\mu\nu}^{ij})_{|Q|\times |Q|})_{(T_M+1)\times (T_M+1)}
\]
Solution grids in compatibility matrix $B$ are in one-to-one relation \ref{e:one-to-one} with the $T_M$-walks in digraph $g$. Let $C=\{c\}$ be the set of all computational threads of machine $M$. Then, $|c| \leq T_M+1$ for any $c\in C$. And each $c\in C$ creates a $(|c|-1)$-walk in digraph $g$. Let's agree to continue each such walk by the $T_M-|c|+1$ loops in the case of $|c|< T_M+1$. And let $w_c$ be that $T_M$-walk in digraph $g$ (possibly ending with the added loops) which was created by computational thread $c\in C$. Then, there are index functions $\phi_c$ in compatibility matrix $B$ which are matched to walks $w_c$ by one-to-one relation \ref{e:one-to-one}. Then, the following compatibility matrix does exist:
\[
B\acute{~} = ((0)_{|Q|\times |Q|})_{(T_M+1)\times (T_M+1)}~\vee~\bigvee_{c\in C}((b_{\phi_c(i)\phi_c(j)}^{ij})_{|Q|\times |Q|})_{(T_M+1)\times (T_M+1)}
\] 
- where $b_{\phi_c(i)\phi_c(j)}^{ij}$ are the appropriate components of our compatibility matrix $B$. And solution grids in compatibility matrix $B\acute{~}$ are in one-to-one relation with the accepting computational threads of machine $M$:
\[
\phi_c~\stackrel{1-1}{\longleftrightarrow}~w_c~\stackrel{1-1}{\longleftrightarrow}~c
\]
- where $c$ is an accepting computational thread.
\begin{theorem}
\label{t:domain}
The compatibility matrix method's domain consists of those Turing-decidable problems whose time-complexity is finite. And any such problem allows an adequate encoding by a compatibility matrix whose size is linear in the problem's time-complexity.
\end{theorem}  
\begin{proof}
We just have seen that any Turing-decidable problem of the finite time-complexity allows such an encoding. And on the other hand, we have seen in the previous sections, any problem which allows the compatibility-matrix encoding can be solved by a decider (Sipser's decider \cite{19}, Kozen's total Turing machine \cite{20}) in the time polynomial in the compatibility matrix's size.
\end{proof}
\begin{xca}
For the given NDTM of the finite time-complexity, we can replace the polynomial time-bond in the Garey and Johnson's proof \cite{21} of Cook-Levin theorem \cite{09,22} by the machine's time-complexity. Then, the argument will produce a SAT instance with a quadratic number of clauses and variables, quadratic in the machine's time-complexity. And compatibility matrix \ref{e:cnf1} for that SAT instance will be another compatibility-matrix encoding of the given NDTM.
\end{xca}
Theorem \ref{t:domain} validates all the above results for the Turing-decidable problems whose time-complexity is bounded by a finite function. And the results may be summarized as the following complements to Savitch's theorem \cite{23}:
\[
\mbox{NTIME}(t) ~\subseteq ~ \mbox{DTIME}(t^8),~\mbox{NSPACE}(s) ~\subseteq ~ \mbox{DTIME}(s^{16})
\]
And it directly proves
\[
\begin{array}{cl}
\mbox{P = NP = PSPACE} ~& \\
\mbox{EXPTIME = NEXPTIME = EXPSPACE} ~& \\
\mbox{$k$-EXPTIME = $k$-NEXPTIME = $k$-EXPSPACE}, & k\geq 2 \\
\end{array} 
\]
- and so on for other complexity classes closed under Karp reduction.
\newline\indent
Compatibility-matrix encoding \ref{e:1ten} is Post reduction of the encoded problem to the following finite recursive language\footnote{"$\circ$" is the concatenation operator.}:
\[
L = [m_1]\circ [m_2]\circ \ldots \circ [m_n]
\]
And any finite recursive language allows the compatibility-matrix encoding due to Theorem \ref{t:domain}. So, domain of the compatibility matrix method may be called FINITE.

\section{Compatibility tensor}
\label{s:ten}

Turing-decidable problems can be encoded with their multidimensional views as well as with their 2D-views - it is often even easier. The multidimensional views will create the multidimensional Boolean box arrays aka \emph{Boolean tensors}. And these blueprints can be formalized as follows.
\begin{definition}
\emph{Compatibility tensor} $T$ of the $\left(_k^k\right)$-valency is a symmetric $k$-dimensional array of the Boolean $k$-dimensional arrays:
\begin{equation}
\label{e:ten}
\left\{\begin{array}{rrcl}
1). & T &=& (T_{i_1i_2\ldots i_k})_{\mbox{\scriptsize $\underbrace{n\times n\times\ldots\times n}_{k}$}} \\
2). & T_{i_1i_2\ldots i_k}&=& (t_{\mu_1\mu_2\ldots\mu_k}^{i_1i_2\ldots i_k})_{m_1\times m_2\times\ldots m_k} \\
3). & t_{\mu_1\mu_2\ldots\mu_k}^{i_1i_2\ldots i_k} &\in & \{false,true\} \\
4). & t_{\ldots \mu \ldots \nu \ldots}^{\ldots i \ldots j \ldots} &=& t_{\ldots \nu \ldots \mu \ldots}^{\ldots j \ldots i \ldots} \\
5). & t_{\mu_1\mu_2\ldots\mu_k}^{i_1i_2\ldots i_k} &\Rightarrow & \bigwedge_{\alpha,\beta\in[k]} (i_\alpha\neq i_\beta~\vee~\mu_\alpha = \mu_\beta)
\end{array}\right.
\end{equation}
- where numbers $n, m_1, m_2, \ldots, m_n$ and maximum $m = \max_{i\in[n]} m_i$ are the tensor's sizes; and number $k$ is the tensor's dimension. The $k$-dimensional boxes\footnote{ Some of the boxes will dimensionally degenerate when $m_i = 1$ for some $i\in[n]$.} $T_{i_1i_2\ldots i_k}$ are called compatibility boxes. The components $t_{\mu_1\mu_2\ldots\mu_k}^{i_1i_2\ldots i_k}$ which equals true are $true$-components, and the rest components of the tensor are  $false$-components.
\newline\indent
The tensor's index function is any one-meaning function
\[
\phi:~i\in[n]~\mapsto~\phi(i)\in[m_i]
\]
Each index function creates the following $k$-dimensional sub-array  of the tensor which is called grid of components or just grid:
\[
\{t_{\phi(i_1)\phi(i_2)\ldots\phi(i_k)}^{i_1 i_2 \ldots i_k}\}_{i_1,i_2,\ldots,i_k\in[n]}
\]
This grid of components is a solution grid if all its components equals $true$, i.e. if its index function satisfies the following functional equation:
\begin{equation}
\label{e:grid}
t_{\phi(i_1)\phi(i_2)\ldots\phi(i_k)}^{i_1 i_2 \ldots i_k} = true,~i_1,i_2,\ldots,i_k\in[n]
\end{equation}
- solution grid in compatibility tensor $T$ is a $k$-dimensional orthogonal lattice of $true$-components, one component per compatibility box. 
\newline\indent
Further, $true$-component $t_{\mu_1^0\mu_2^0\ldots\mu_k^0}^{i_1^0i_2^0\ldots i_k^0} = true$ is a noise if functional equation \ref{e:grid} is inconsistent subject to the following constrains
\[
\phi(i_1^0) = \mu_1^0,~\phi(i_2^0) = \mu_2^0,~\ldots,~\phi(i_k^0) = \mu_k^0
\]
Otherwise, $true$-component $t_{\mu_1^0\mu_2^0\ldots\mu_k^0}^{i_1^0i_2^0\ldots i_k^0}$ belongs to a solution grid.
\newline\indent
Inversion of all noisy components in compatibility tensor will transform it into its general solution. The general solution is a compatibility tensor on its own. And the general solution of tensor \ref{e:ten} contains $true$-components iff compatibility tensor \ref{e:ten} contains solution grids.
\end{definition}
\begin{xca}
Any compatibility matrix \ref{e:1ten} is a compatibility tensor of the $\left(_2^2\right)$-valency; the matrix's major diagonal is a compatibility tensor of the $\left(_1^1\right)$-valency; and for any $k\geq 1$, the following conjunctions create a compatibility tensor of the $\left(_k^k\right)$-valency:
\[
t_{\mu_1\mu_2\ldots\mu_k}^{i_1i_2\ldots i_k} = \bigwedge_{(i,\mu),(j,\nu)\in\{(i_1,\mu_1),(i_2,\mu_2),\ldots,(i_k,\mu_k)\}} b_{\mu\nu}^{ij}
\]
- where $i_1,i_2,\ldots,i_k\in[n]$ and $(\mu_1,\mu_2,\ldots,\mu_k)\in[m_{i_1}]\times[m_{i_2}]\times\ldots\times[m_{i_k}]$.
\end{xca}
Compatibility tensors are used for encoding of problems exactly as the compatibility matrices. And the only requirement is the encoding adequacy: there are solution grids in the \emph{compatibility-tensor encoding} of the given problem iff that problem is consistent, i.e. iff it has solutions.
\begin{xca}
Let the compatibility relations in system \ref{e:ro} be $k$-ary: 
\[
\rho_{i_1i_2\ldots i_k} \subseteq O_{i_1}\times O_{i_2}\times\ldots\times O_{i_k}
\]
- where (partially) diagonal relations are the appropriate (partially) inclusion relations. Then, the $k$-dimensional graphics of these relations will create an adequate compatibility-tensor encoding of the system. 
\end{xca}
Everything from the previous sections can be edited for the compatibility tensors. Yet, let's notice, the solution grids existence problem for compatibility tensor \ref{e:ten} is Turing-decidable in the finite time: we just iterate all $\prod_{i\in[n]}m_i$ index functions of the tensor and test them with equation \ref{e:grid}. Therefore, due to Theorem \ref{t:domain}, any compatibility tensor allows an adequate compatibility-matrix encoding. It can be done with the following protocol: \emph{in the compatibility matrix of a compatibility tensor, the matrix's off-diagonal compatibility boxes encode the tensor's box structure while the matrix's diagonal compatibility boxes encode the tensor's components.}
\begin{xca}
For compatibility tensor \ref{e:ten}, the protocol can be realized as follows (in the enumeration invariant form):
\begin{equation}
\label{e:bt}
B = (B_{\alpha_1\alpha_2})_{n^k\times n^k} = ((b_{\beta_1\beta_2}^{\alpha_1\alpha_2})_{\prod_{\omega=1}^k m_{i_\omega^1}\times \prod_{\omega=1}^k m_{i_\omega^2}})_{n^k\times n^k}
\end{equation}
- where 
\[
\begin{array}{l}
\alpha_1 = (i_1^1,i_2^1,\ldots,i_k^1)\in [n]^k; \\
\alpha_2 = (i_1^2,i_2^2,\ldots,i_k^2)\in [n]^k; \\
\beta_1 = (\mu_1^1,\mu_2^1,\ldots,\mu_k^1)\in [m_{i_1^1}]\times [m_{i_2^1}]\times \ldots \times [m_{i_k^1}]; \\ 
\beta_2 = (\mu_1^2,\mu_2^2,\ldots,\mu_k^2)\in [m_{i_1^2}]\times [m_{i_2^2}]\times \ldots \times [m_{i_k^2}]; \\
\end{array} 
\]
and
\[
b_{\beta_1\beta_2}^{\alpha_1\alpha_2} = (\alpha_1\neq\alpha_2 \vee t_{\beta_1}^{\alpha_1})~\wedge~(\alpha_1=\alpha_2 \vee \bigwedge_{\omega_1,\omega_2\in [k]} (i_{\omega_1}^1\neq i_{\omega_2}^2~\vee~\mu_{i_{\omega_1}^1} = \mu_{i_{\omega_2}^2}));
\]
where $t_{\beta_1}^{\alpha_1} = t_{\mu_1^1\mu_2^1\ldots \mu_k^1}^{i_1^1i_2^1\ldots i_k^1}$ are the tensor's components. 
\newline\indent
For example, compatibility matrix \ref{e:rk} is a rationalization of compatibility matrix \ref{e:bt} for compatibility tensor \ref{e:1ten}.
\end{xca}

\section*{Conclusion}

This article presented the compatibility matrix method, a generalization of descriptive geometry on the combinatorial problems. In the method, problems are encoded by the "blueprints" of their solutions, and the solutions themselves become an orthogonal-lattice pattern in the encoding. Such a regularization of the problems allows perform parallel testing of all guesses with the efficient deterministic algorithms - the article presented a few. The encoding is Post reduction of Turing-decidable problems to finite recursive languages, and computational complexity of the compatibility matrix method is basically polynomial in the computational complexity of that reduction. Further, domain of the compatibility matrix method is exactly the class of finite recursive languages. The class was called FINITE. And, because any Rado's busy beaver \cite{24} belongs to FINITE (it is easy to see),
\[
\mbox{ELEMENTARY $\subset$ FINITE $\subseteq$ R}
\]
And, the compatibility matrix method shows, the time and space complexities of the problems in class FINITE are polynomially equivalent.
\newline\indent
There are some demos of the compatibility matrix method at \cite{25}.


\begin{thebibliography}{99}
\bibitem{01} Sergey Gubin, {\it Complementary to Yannakakis' theorem,} JCMCC 74 (2010), pp. 313-321 (see arXiv:cs/0610042v3 [cs.DM])
\bibitem{02} Sergey Gubin, {A Sudoku solver,} Proceedings of the World Congress on Engineering and Computer Science 2009 Vol I WCECS 2009, October 20-22, 2009, San Francisco, USA, ISBN: 978-98817012-6-8, pp. 223-228
\bibitem{03} Sergey Gubin, {\it Polynomial size asymmetric linear model for SAT}, In Proc. of the Advances in Electrical and Electronics Engineering - IAENG Special Edition of 
the World Congress on Engineering and Computer Science 2008 (WCECS 2008), ISBN-13: 978-0-7695-3555-5, pp. 62-66 \url{http://doi.ieeecomputersociety.org/10.1109/WCECS.2008.16}
\bibitem{04} Sergey Gubin, {\it Polynomial size asymmetric linear model for Subgraph Isomorphism},
Proceedings of the World Congress on Engineering and Computer Science 2009, WCECS 2008, ISBN: 978-988-98671-0-2, pp. 241-246 (see arXiv:0802.2612v2 [cs.DM])
\bibitem{05} Sergey Gubin, {\it Compatibility matrix method,} 25-th MCCCC, Program and Abstracts, p. 9, University of Nevada, Las Vegas, Nevada, 2011
\bibitem{06} Sergey Gubin, {\it On the purely logical solution of Sudoku,} 24-th MCCCC, Program and Abstracts, p. 9, Illinois State University, Normal, Illinois, 2010
\bibitem{07} Sergey Gubin, {\it A lexicographical approach to SAT,} 23-nd MCCCC, Program and Abstracts, p. 9, Rochester Institute of Technology,  Rochester, New York, 2009
\bibitem{08} Stephen Cook, {\it The P versus NP Problem,}
\url{http://www.claymath.org/millennium/P_vs_NP/Official_Problem_Description.pdf}
\bibitem{09} Stephen Cook, {The complexity of theorem-proving procedures.} In Conference Record of Third Annual ACM Symposium on Theory of Computing, pages 151-158, 1971
\bibitem{10} Richard M. Karp, {\it Reducibility among combinatorial problems.} In R. E. Miller and J. W. Thatcher, editors, Complexity of Computer Computations, pages 85-103, New York, 1972. Plenum Press
\bibitem{29} Karp, R. M., Lipton, R. J. (1980), {\it Some connections between nonuniform and uniform complexity classes,} Proceedings of the Twelfth Annual ACM Symposium on Theory of Computing, pp. 302-309
\bibitem{11} Schaefer, Thomas J. (1978), {\it The Complexity of Satisfiability Problems.} STOC 1978. pp. 216-226
\bibitem{12} Mihalis Yannakakis, {\it Expressing combinatorial optimization problems by linear programs,} In Proc. of the twentieth annual ACM Sympos. on Theory of computing, Chicago, Illinois, pp.
223-228, 1988
\bibitem{28} Rossman, Benjamin (2008), {\it On the constant-depth complexity of k-clique,} STOC 2008: Proceedings of the 40th annual ACM symposium on Theory of computing. ACM. pp. 721-730
\bibitem{13} Grover L.K., {\it A fast quantum mechanical algorithm for database search,} Proceedings, 28th Annual ACM Symposium on the Theory of Computing, (May 1996) p. 212 
\bibitem{14} T. Baker, J. Gill, and R. Solovay, {\it Relativizations of the P =? NP question.} SICOMP:  SIAM Journal on Computing, 1975
\bibitem{15} S. Fortune, {\it A note on sparse complete sets.} SIAM Journal on Computing, volume 8, issue 3, pp.431-433. 1979
\bibitem{16} S. R. Mahaney, {\it Sparse complete sets for NP: Solution of a conjecture by Berman and Hartmanis.} Journal of Computer and System Sciences 25:130-143. 1982
\bibitem{31} Ravi Kannan. {\it Circuit-size lower bounds and non-reducibility to
sparse sets,} Information and Control, 55:40-56, 1982
\bibitem{17} Alexander A. Razborov, {\it Lower bounds on the monotone complexity of some boolean functions.} Soviet Math. Dokl., 31:354-357, 1985
\bibitem{18} N. Alon and R. B. Boppana, {\it The monotone circuit complexity of boolean functions.} Combinatorica, 7(1):1-22, 1987
\bibitem{26} Chandra, A.K., and Stockmeyer, L.J, {\it Alternation,} Proc. 17th IEEE Symp. on Foundations of Computer Science, Houston, Texas, 1976, pp. 98-108
\bibitem{27} Chandra, A.K. and Kozen, D.C. and Stockmeyer, L.J., {\it Alternation,} Journal of the ACM, Volume 28, Issue 1, pp. 114-133, 1981. 
\bibitem{19} Sipser, M. (1996), {\it Introduction to the Theory of Computation,} PWS Publishing Co.
\bibitem{20} Kozen, D.C. (1997), {\it Automata and Computability,} Springer
\bibitem{21} M. R. Garey and D. S. Johnson, {\it Computers and Intractability, a Guide to the Theory of NP-Completeness.} W.H. Freeman and Co., San Francisco, 1979
\bibitem{22} L. Levin. {\it Universal search problems (in Russian).} Problemy Peredachi Informatsii, 9(3):265-266, 1973. English translation in Trakhtenbrot, B. A.: {\it A survey of Russian approaches to Perebor (brute-force search) algorithms.} Annals of the History of Computing, 6 (1984), pages. 384-400
\bibitem{23} Savitch, Walter J. (1970), {\it Relationships between nondeterministic and deterministic tape complexities}, Journal of Computer and System Sciences 4 (2):177-192
\bibitem{24} Rad$\acute{\mbox{o}}$, Tibor (1962), On non-computable functions, Bell System Technical Journal, Vol. 41, No. 3 (May 1962), pp. 877-884
\bibitem{25} Polynomial times, \url{http://www.timescube.com}
\end{thebibliography}
\end{document}